\documentclass[preprint,12pt]{elsarticle}

\usepackage{amssymb}
\usepackage{subcaption}
\usepackage{url} 
\usepackage{upgreek}
\usepackage{xcolor}
\usepackage{amsmath}

\usepackage{lineno}
\journal{NIM A}

\begin{document}

\begin{frontmatter}



\title{A Comparison of Micromegas with $x/y$ Strip Charge Readouts for Directional Recoil Detection}

\author[UH]{Majd Ghrear \corref{cor1}}
\author[UoS]{Alasdair G. McLean}
\author[UH]{Hima B. Korandla}
\author[ANU,ARC]{Ferdos Dastgiri}
\author[UoS]{Neil J. C. Spooner}
\author[UH]{Sven E. Vahsen}

\cortext[cor1]{Corresponding author, email address: majd@hawaii.edu}
\affiliation[UH]{organization={Department of Physics and Astronomy, University of Hawaii},%Department and Organization
            city={Honolulu},
            postcode={96822}, 
            state={HI},
            country={USA}}
            
\affiliation[UoS]{organization={Department of Physics and Astronomy, University of Sheffield},%Department and Organization
            city={Sheffield},
            postcode={S3~7RH}, 
            country={UK}}

\affiliation[ANU]{organization={Department of Nuclear Physics and Accelerator Applications, Research School of Physics, Australian National University},%Department and Organization
            city={Canberra},
            postcode={ACT 2601}, 
            country={Australia}}

\affiliation[ARC]{organization={ARC Centre of Excellence for Dark Matter Particle Physics},%Department and Organization
            country={Australia}}

\begin{abstract}
Detecting the topology and direction of low-energy nuclear and electronic recoils is broadly desirable in nuclear and particle physics, with applications in coherent elastic neutrino-nucleus scattering (CE$\nu$NS), astrophysical neutrino measurements, probing dark matter (DM) beneath the neutrino fog, and confirming the galactic origin of DM. Gaseous Time Projection Chambers (TPCs) offer the required gain and readout granularity, but must be large to achieve the required volume. Therefore, scalable, cost-effective TPC readout technologies are essential. High-resolution $x/y$ strip readouts, previously identified as the optimal balance between cost-efficiency and performance, are examined here. To guide the readout design of a 40-L detector under construction, we present a comparative analysis of nine $x/y$ strip configurations with Micromegas amplification. Each setup employs VMM3a front-end ASICs within the RD51 Scalable Readout System (SRS) for strip readout and a pulse height analyzer for reading out the Micromegas mesh. These complementary techniques assess gain, gain resolution, $x/y$ charge sharing, and spatial resolution of each setup. Configurations with a diamond-like carbon (DLC) layer exhibit improved spark resistance, allowing larger maximal gain and improved fractional gain resolution without notable impact on the spatial resolution. Although the DLC reduces the signal in the strips situated lower in the readout plane, this can be mitigated by narrowing the perpendicularly oriented strips above them. Our results allow us to select the optimal readout for future detectors. We also observe clear 3D tracks from alpha particles, with performance in good agreement with a simple simulation. Overall, Micromegas with $x/y$ strip readout are promising for low-energy recoil observatories. However,  dedicated amplification devices and/or improved electronics are needed to reach the fundamental performance limit of 3D electron counting.

\end{abstract}


% \begin{highlights}
% \end{highlights}


\end{frontmatter}


\section{Introduction}
\label{intro}

Over the past three decades, direct detection efforts for conventional weakly interacting massive particle (WIMP) dark matter (DM) have made substantial progress, ruling out large regions of WIMP parameter space~\cite{Cooley:2022ufh,Schumann:2019eaa,Billard:2021uyg}. Non-gravitational evidence of galactic DM interacting with Standard Model (SM) particles remains absent. As detectors improve sensitivity to probe lower masses and smaller cross sections, the once negligible neutrino background becomes increasingly significant, eventually overshadowing potential WIMP signals. In fact, some direct detection experiments have recently reported measurements of this background~\cite{PandaX:2024muv,XENON:2024ijk}. Known as the neutrino fog,  it presents a difficult obstacle for conventional DM detectors that cannot differentiate it from a DM signal~\cite{OHare:2021utq}. This challenge has sparked renewed interest in directional dark matter detection.

Directional dark matter detection was first proposed in Ref.~\cite{PhysRevD.37.1353} which recognized that DM-induced nuclear recoils are subject to a unique directional signature caused by the motion of our solar system with respect to our galaxy's DM halo. Modern gaseous time projection chambers (TPCs) are uniquely capable of reconstructing the directions of such recoils~\cite{Vahsen:2021gnb}. This opens the possibility of a new generation of dark matter experiments capable of circumventing the neutrino fog and also confirming the galactic origin of a DM signal.

Directional recoil detection, the general ability to detect the direction of nuclear and electron recoils, also has interesting applications beyond dark matter. The COHERENT collaboration has now detected coherent elastic neutrino-nucleus scattering (CEvNS) in CsI~\cite{CsI}, Ar~\cite{Ar}, and Ge~\cite{Ge}. While current experiments only provide information about the nuclear recoil energy, additional information about the direction of the recoil can be valuable in discerning new physics~\cite{DCEvNS1,DCEvNS2}. Directional detectors can also be deployed for studying solar neutrinos~\cite{OHare:2022jnx}. Low-energy solar neutrino fluxes such as $pp$, $pep$, $^7Be$, and $CNO$ could be accessed via neutrino-electron elastic scattering. Ref.~\cite{Lisotti:2024fco} finds that a large gas-based directional recoil observatory intended for dark matter searches can double as a competitive directional neutrino experiment. 

In a study on the feasibility of a large-scale directional recoil observatory with sensitivity to both dark matter and neutrinos~\cite{Vahsen:2020pzb}, gas TPCs with high-resolution $x/y$ strip readout were identified as the optimal trade-off between cost and performance. To optimize the design of a 40-L prototype detector being constructed by our group, we perform an experimental comparative analysis of nine different, highly segmented $x/y$ strip TPC charge readout plane configurations. All configurations utilize a bulk Micromegas amplification structure~\cite{GIOMATARIS199629,GIOMATARIS2006405} and are tested in a common, miniature TPC. We compare the configurations by assessing their gain, gain resolution, $x/y$ strip charge sharing, and spatial resolution. In Section~\ref{exp-setup}, we detail the experimental setup. In Section~\ref{PHA}, we characterize the readouts by analyzing the pulses of charge drawn by the Micromegas mesh. In Section~\ref{VMM}, we characterize the readouts by reading out their $x/y$ strips using VMM3a front-end ASICs~\cite{Iakovidis_2020} within the RD51 SRS~\cite{Scharenberg_2022,PFEIFFER2022166548,LUPBERGER201891}. An algorithm is developed to reconstruct the digital VMM3a output data into 3D tracks. The algorithm is demonstrated on alpha tracks, which are then used to assess the spatial resolution of the configurations. Section~\ref{Discussion} combines the findings from Sections~\ref{PHA} and~\ref{VMM} to comprehensively compare the readout configurations and discuss future directions.

\section{Experimental Setup}
\label{exp-setup}

\begin{figure}[ht]
\begin{center}
\includegraphics[width=\textwidth,trim={0cm .2cm 0cm 0.2cm},clip]{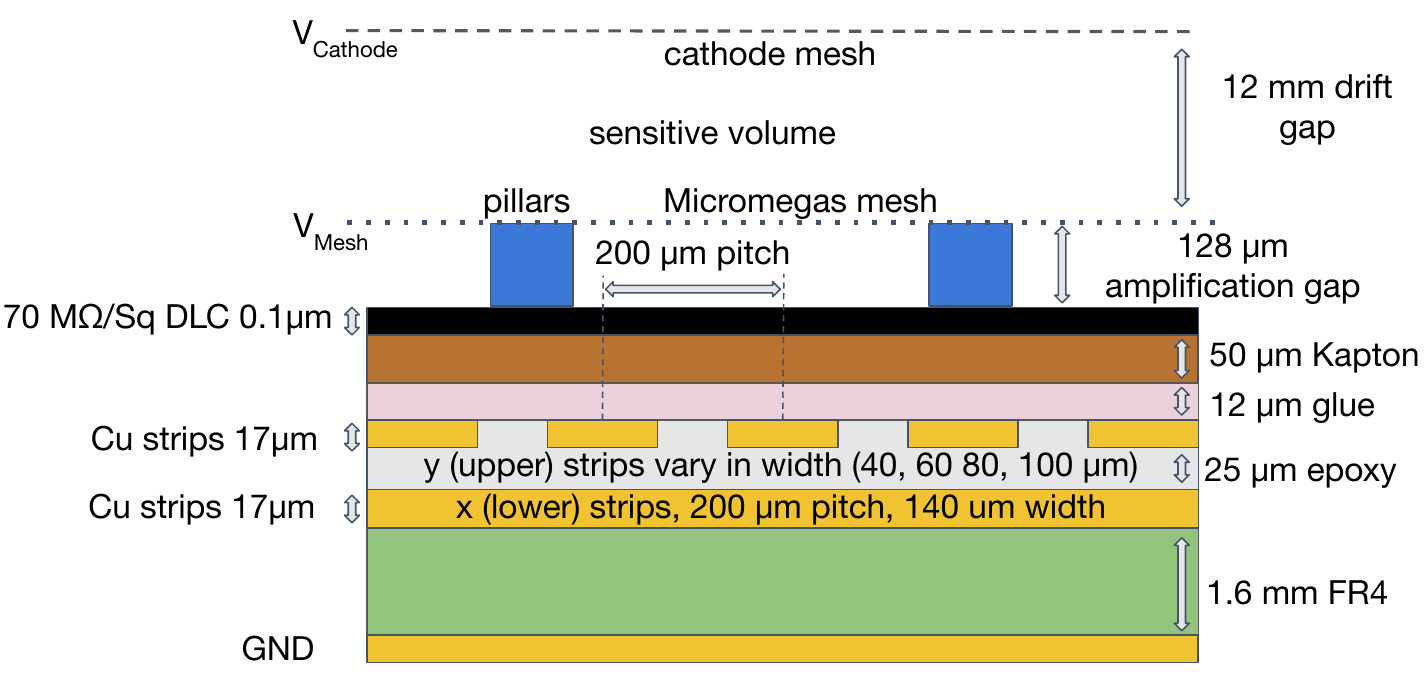}
%\framebox[4.0in]{$\;$}
%\fbox{\rule[-.5cm]{0cm}{4cm} \rule[-.5cm]{4cm}{0cm}}
\end{center}
\caption{Cross-sectional view of the UH DLC readout plane. The pitch of the $x$ (lower) strips and $y$ (upper) strips is uniformly 200\,$\upmu$m. The width of the $y$ (upper) strips varies with each quadrant, as depicted in Figure~\ref{top_view}. This figure is not to scale.}
\label{cross_sect}
\end{figure}

\begin{figure}[ht]
\begin{center}
\includegraphics[width=.9\textwidth,trim={.5cm 0cm 0.1cm 0cm},clip]{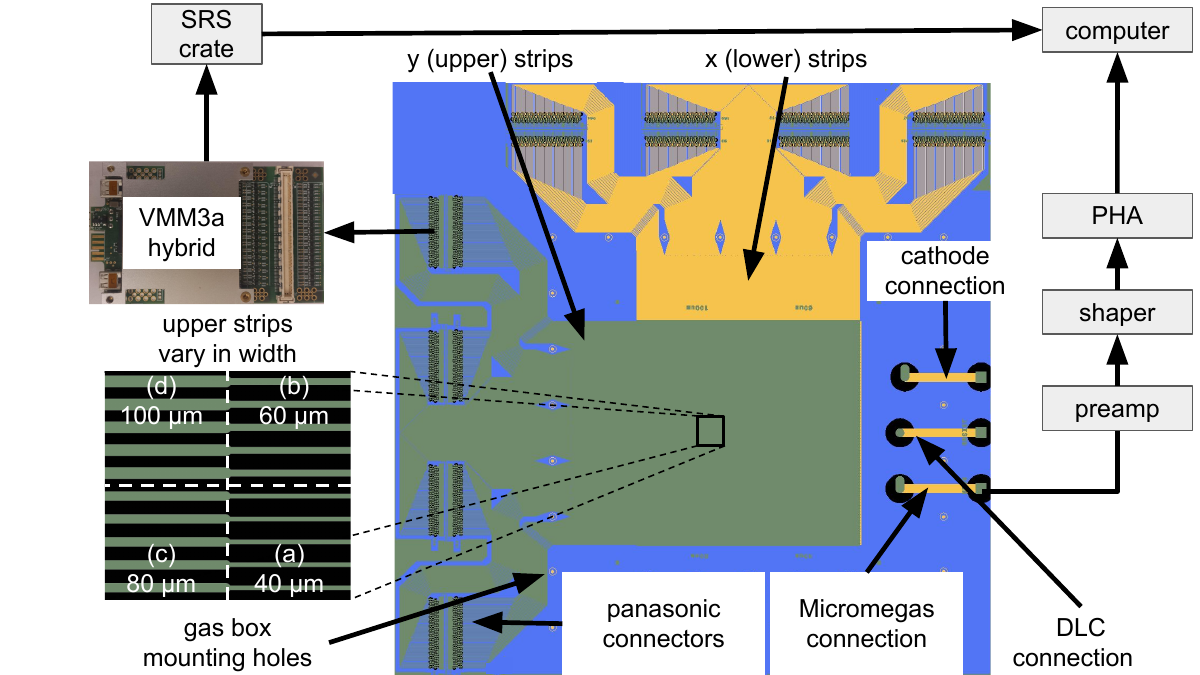}
%\framebox[4.0in]{$\;$}
%\fbox{\rule[-.5cm]{0cm}{4cm} \rule[-.5cm]{4cm}{0cm}}
\end{center}
\caption{A top view of the UH DLC readout plane. All readout planes have compatible mounting holes so that the same gas box, spacers, and cathode mesh can be attached. The Micromegas amplification structure, which sits on top of the readout, is not displayed. An enlarged view of the quadrants, denoted $a$, $b$, $c$, and $d$, shows the four unique configurations with varying width of the $y$ (upper) strips. The lower strip width is 140\,$\upmu$m and does not vary between quadrants. The strip charge signal is amplified and digitized with the VMM3a front-end ASIC within the RD51 Scalable Readout System. The Micromegas mesh avalanche charge signal is amplified and read out with a charge-sensitive preamplifier, followed by a shaping amplifier and pulse height analyzer.}
\label{top_view}
\end{figure}

Our experimental setup utilizes three charge readout planes. The first readout plane is procured by the University of Hawaii (UH) and is coated with a diamond-like carbon (DLC) layer~\cite{ROBERTSON2002129,LV2020162759} with a resistivity of 70\,M$\Omega$/Sq, therefore we denote it as `UH DLC'. A cross-sectional view of the UH DLC readout plane is depicted in Figure~\ref{cross_sect} and a top view is presented in Figure~\ref{top_view}. A unique aspect of UH DLC, illustrated by the inset of Figure~\ref{top_view}, is that it is divided into four quadrants, each with varying upper strip widths. This means that the single readout plane encompasses four distinct $x/y$ strip configurations. The second readout plane, also procured by UH, does not include a DLC layer and is therefore denoted `UH NoDLC'. UH NoDLC mirrors UH DLC in design except that the DLC, glue and Kapton layers are omitted (see Figure~\ref{cross_sect}). The third readout plane is procured by the University of Sheffield and therefore is denoted `UoS'. The UoS readout plane is coated with a 50\,M$\Omega$/Sq DLC layer. Unlike the UH configurations, the UoS plane is not divided into different quadrants. All of the readout plane have a $10$\,cm$ \times 10$\,cm readout area. The specifications of the readout planes are summarized in Table~\ref{MM_config}. The UH DLC, UH NoDLC, and UoS readout plane collectively encompass nine distinct $x/y$ strip configurations.

\begin{table}[h!]
\centering
\begin{tabular}{|c| c c c|} 
 \hline
 Detector Name & UH DLC & UH NoDLC & UoS \\ [0.5ex] 
 \hline
 Amplification gap [$\upmu$m] & 128 & 128 & 256 \\ 
 DLC Resistivity [M$\Omega$/Sq] & 70 & N/A & 50 \\ 
Strip Pitch [$\upmu$m] & 200 & 200 & 250 \\
 Quadrant Names & $a,b,c,d$ & $a,b,c,d$ & N/A \\
 $y$ (upper) strip width [$\upmu$m] & 40, 60, 80, 100 & 40, 60, 80, 100 & 100 \\
 $x$ (lower) strip width [$\upmu$m] & 140 & 140 & 220 \\ [1ex] 
 \hline
\end{tabular}
\caption{Specifications of the readout planes under test. The two UH readout planes are split into four quadrants: $a$, $b$, $c$, and $d$, with varying $y$ (upper) strips width: 40, 60, 80, and 100\,$\upmu$m, respectively, as illustrated in Figure~\ref{top_view}. All readout planes utilize a bulk Micromegas amplification structure.}
\label{MM_config}
\end{table}

\begin{figure}
\centering
\begin{subfigure}[b]{.6\linewidth}
\includegraphics[width=\linewidth]{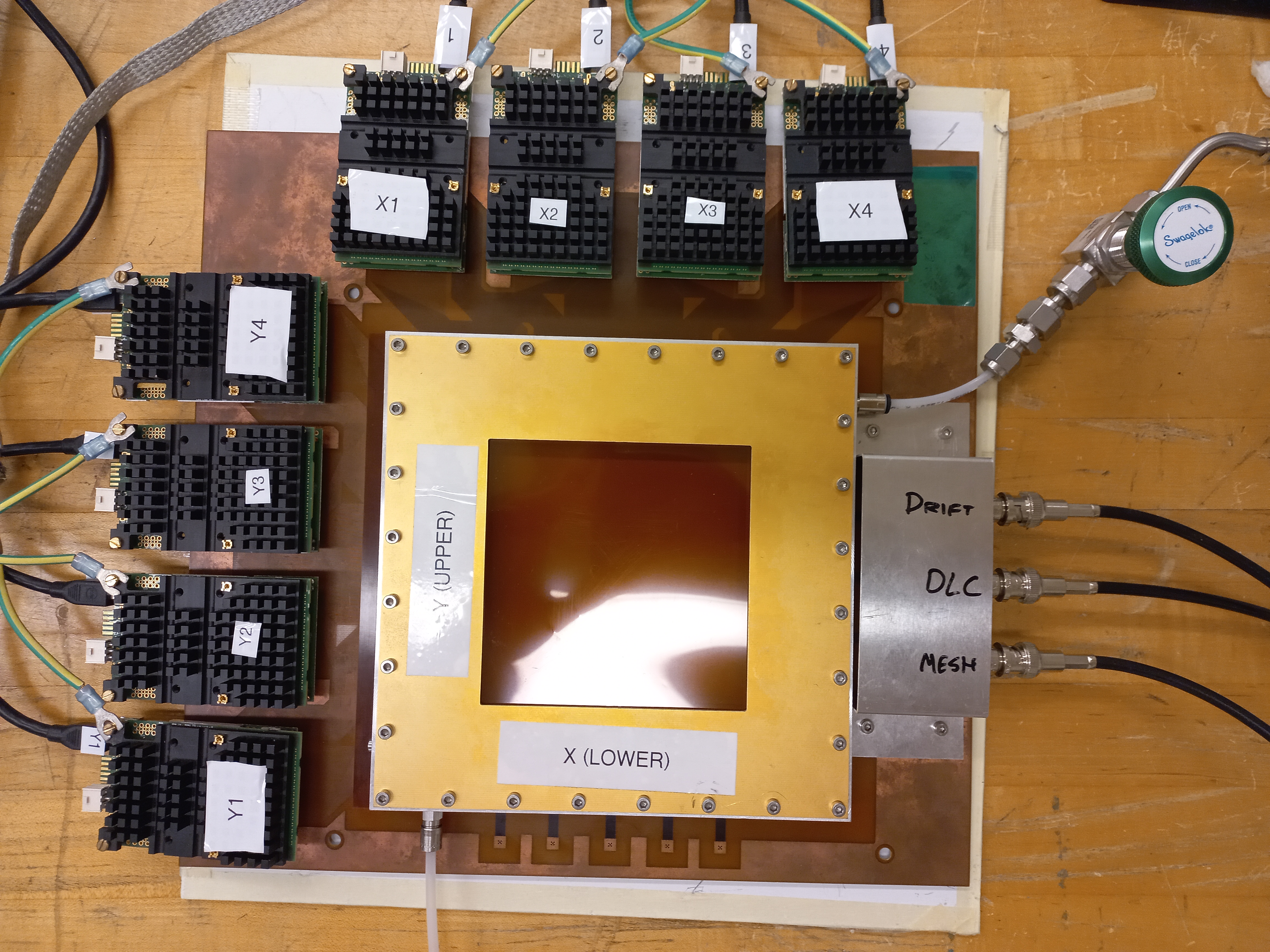}
\caption{}\label{MM_pic}
\end{subfigure}

\begin{subfigure}[b]{.3\linewidth}
\includegraphics[width=\linewidth]{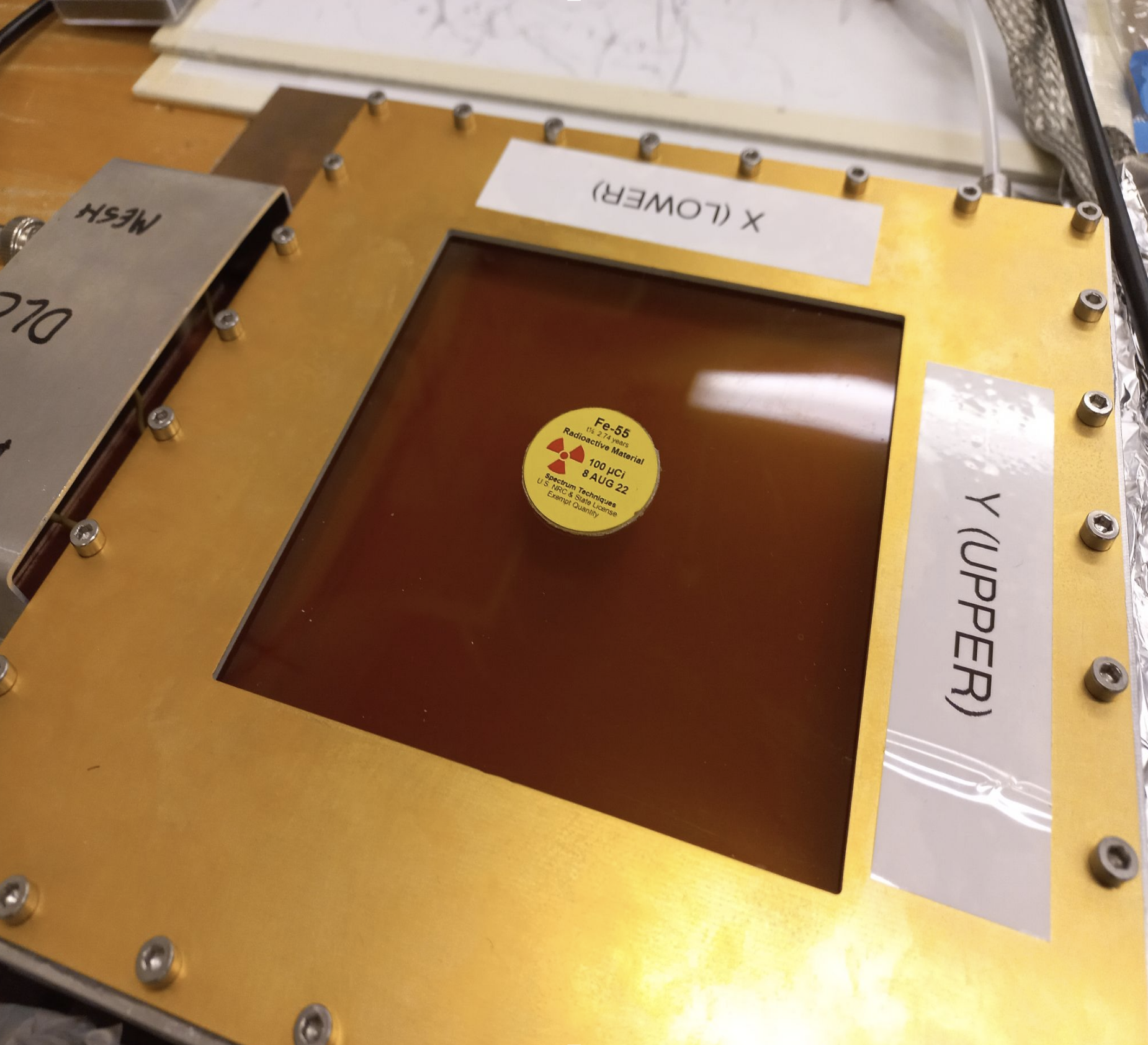}
\caption{}\label{MM_Fe55}
\end{subfigure}
\begin{subfigure}[b]{.3\linewidth}
\includegraphics[width=\linewidth,trim={0cm 4.6cm 0cm 2cm},clip]{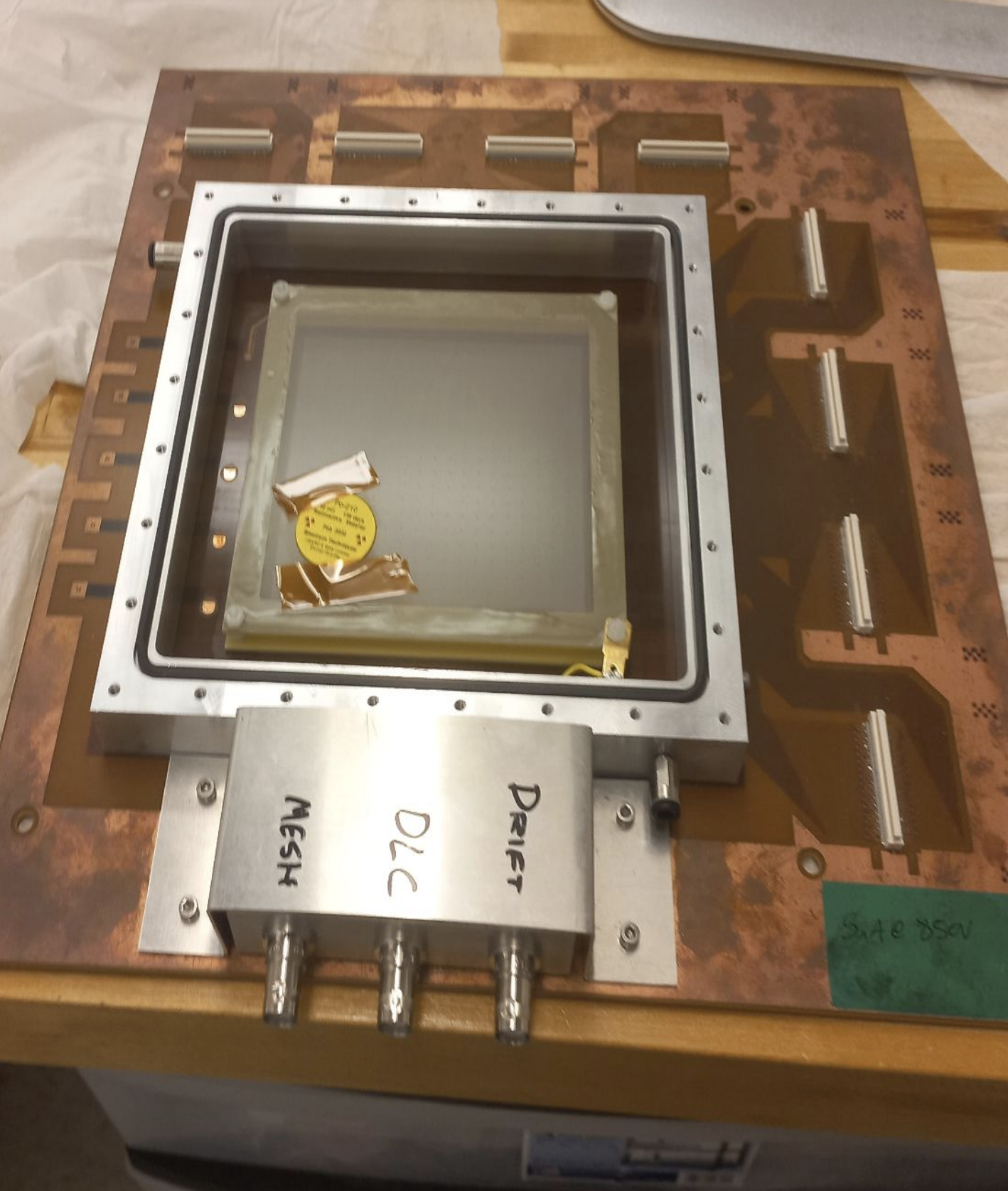}
\caption{}\label{MM_Po210}
\end{subfigure}
\caption{The experimental setup at various stages. (a) The UH DLC detector inside the gas box. VMM front end hybrids are connected to the strips on the $x$ and $y$ axis. (b) An Fe-55 source is placed on the Kapton film above the readout plane, illuminating the sensitive volume with 5.9 keV X-rays. (c) A Po-210 source is placed inside the gas box, directly on top of the cathode mesh.}
\label{sources}
\end{figure}

In a Micromegas TPC, the sensitive volume is situated in the drift gap between the cathode mesh and the Micromegas mesh, illustrated in Figure~\ref{cross_sect}. For all setups, the drift length is 12\,mm and the cathode voltage ($V_\textrm{Cathode}$) is set 504\,V below the Micromegas mesh voltage ($V_\textrm{Mesh}$), so that the drift gap holds a uniform $420$\,V/cm electric (drift) field, corresponding to a drift speed of $8$\,$\upmu$m/ns. Ionizing radiation creates free electrons in the sensitive volume and the uniform field causes them to drift towards the amplification gap. In the amplification gap, a strong electric field is created between the Micromegas mesh which is held at $V_\textrm{Mesh}$ and the DLC layer which is grounded. In the special case of the UH NoDLC readout plane, the strong electric field is created between the Micromegas mesh and the strips, which are also grounded. The strong electric field causes the free electrons to avalanche multiply. The avalanche then induces a charge signal on the $x/y$ strips. The setups are enclosed in a gas box, displayed in Figure~\ref{sources}. We utilize a gas mixture of 70\% He and 30\% $\rm{CO}_2$ maintained at atmospheric pressure and 20$^\circ$C, as used in~\cite{Jaegle:2019jpx}.  

The readout planes are instrumented using two methodologies. One method aims to characterize the avalanche gain and its resolution by measuring the pulses of charge drawn by the Micromegas mesh when the TPC is exposed to an Fe-55 X-ray source. This is done by biasing the Micromegas mesh through a CREMAT CR-150 circuit board~\cite{CR150}. The board includes a CREMAT CR-111 charge sensitive preamplifier whose output is connected to a CR-200-4$\upmu$s shaper module on a CR-160 shaper evaluation board~\cite{CR160}. This approach aligns with the technique previously implemented in Refs.~\cite{Eldridge_2023,McLean:2023dnh}. The output of the shaper module is then connected to an Ortec EASY-MCA Pulse Height Analyzer (PHA) and a computer with the MAESTRO software package~\cite{maestro}, as done in Refs.~\cite{Vahsen:2014fba, Thorpe:2021qce}. Henceforth, we refer to this as the PHA setup. The second method reads out all strips individually using RD51 VMM3a front-end hybrids connected to an SRS data acquisition system (DAQ)~\cite{Scharenberg_2022,PFEIFFER2022166548,LUPBERGER201891}. Each hybrid employs two 64-channel VMM3a ASICs~\cite{Iakovidis_2020}, herein referred to as VMMs, enabling it to read out 128 channels. The integration of the UH DLC detector with these two readout methods is depicted in Figure~\ref{top_view}.

\section{Characterization via Micromegas Mesh Charge Pulses}
\label{PHA}
The top of the gas box consists of a thin Kapton layer directly above the sensitive volume. By placing an uncollimated Fe-55 source above the Kapton, as shown in Figure~\ref{MM_Fe55}, we induce 5.9\,keV X-ray conversion events in the sensitive volume that can be used to measure the avalanche gain and gain resolution of all detectors. To this end, the PHA setup described in Section~\ref{exp-setup} is used. Using a function generator, test pulses are injected into the preamp's test input to obtain a sensitivity of 2.11 PHA bins/fC. Based on repeated calibrations, we estimate this sensitivity to be stable within 2.2\% for the measurements reported here, allowing precise relative comparisons of different detector configurations. We do not attempt a precise absolute calibration of the gain, hence all measurements reported are implicitly subject to an absolute gain uncertainty (due to effects such as gas quality, temperature, pressure, and HV supply calibration uncertainties) of order 25\%. The expected number of primary electrons for a 5.9\,keV X-ray conversion events in our gas mixture is simulated with \texttt{Degrad}~\cite{degrad} as $\textrm{N}_{exp} = 167.5$. The gain of each event ($x$) is obtained by dividing the observed charge by $\textrm{N}_{exp}$. The gain of the photoelectric events is measured over a period of one minute and the gain distribution is fitted to the Crystal Ball function~\cite{CB1,CB2} 
\begin{equation}
        f_{\rm{CB}}(x)=
        \begin{cases}
            \frac{N}{\sigma}e^{-\frac{1}{2}(\frac{x-\mu}{\sigma})^2} , & \textrm{for } x > \mu - \beta \sigma \\
            \frac{N}{\sigma} ( \frac{m}{|\beta|})^m (\frac{m}{|\beta|} - |\beta| - \frac{x - \mu}{\sigma} )^{-m} & \textrm{for } x \leq \mu - \beta \sigma 
        \end{cases}
        \label{crystalball}
    \end{equation}
where $\beta > 0$, $m > 1$, $\mu$, and $\sigma >0$ are fit parameters. The point where the probability density function switches from a power-law to a Gaussian is defined by $\beta$ and $m$ is the power of the tail. The mean and standard deviation of the Gaussian are given by $\mu$ and $\sigma$, respectively. Hence, the avalanche gain as measured on the Micromegas mesh by the PHA setup is denoted by $G$ and obtained as the fit value of $\mu$. Similarly the avalanche gain resolution is denoted by $\sigma_\textrm{G}$ and obtained as the fit value of $\sigma$. For PHA data only, we also include a second-order polynomial background component in the fitting function.

To ensure stable operation, gas is flowed through the gas box, which has a volume of 0.45\,L, at 0.15\,SLPM. This is done with the Fe-55 source placed above the center of the readout so that the gain can be measured versus purge time for each detector. In all cases the gain reaches over 99\% stability within 30 minutes. Furthermore, we confirm that there is no notable increase in $G$ when the gas flow is increased to 0.225\,SLPM, indicating that the flow rate of 0.15\,SLPM is sufficiently high. All data is taken after 30 minutes of gas flow at 0.15\,SLPM.

\subsection{Results}
\label{PHA_results}

\begin{figure}[ht]
\begin{center}
\includegraphics[width=.6\textwidth,trim={.1cm .1cm 1.5cm 1cm},clip]{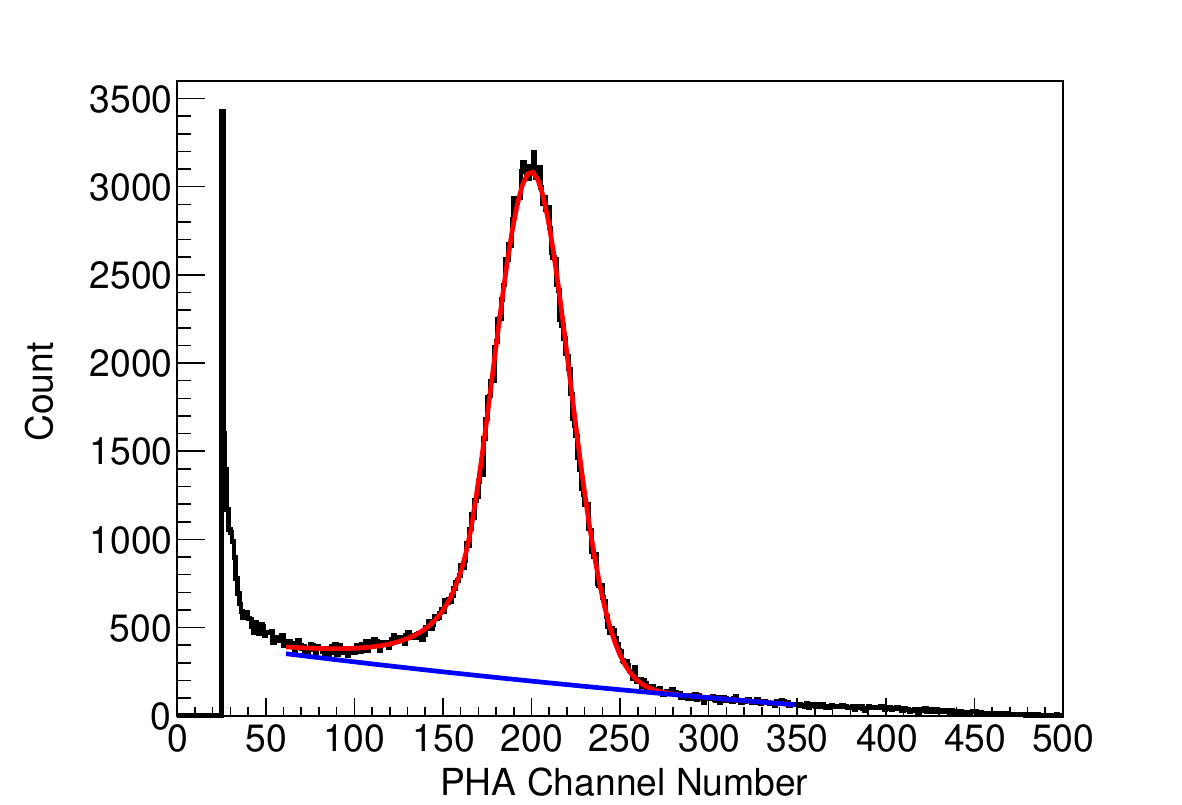}\end{center}
\caption{Example pulse-height distribution measured with a PHA connected to the Micromegas mesh of the UH DLC detector operating at $V_\textrm{Mesh} = 630$\,V. The data are fit to the sum of a Crystal Ball function (red curve) and a second order polynomial (blue curve). The narrow peak at lowest pulse-height is due to noise from the PHA system.}
\label{PHAdist}
\end{figure}

\begin{figure}[ht]
    \begin{subfigure}{.495\textwidth} 
        \centering
        % include first image
        \includegraphics[width=\linewidth,trim={.3cm 0.4cm 0.35cm 0.35cm},clip]{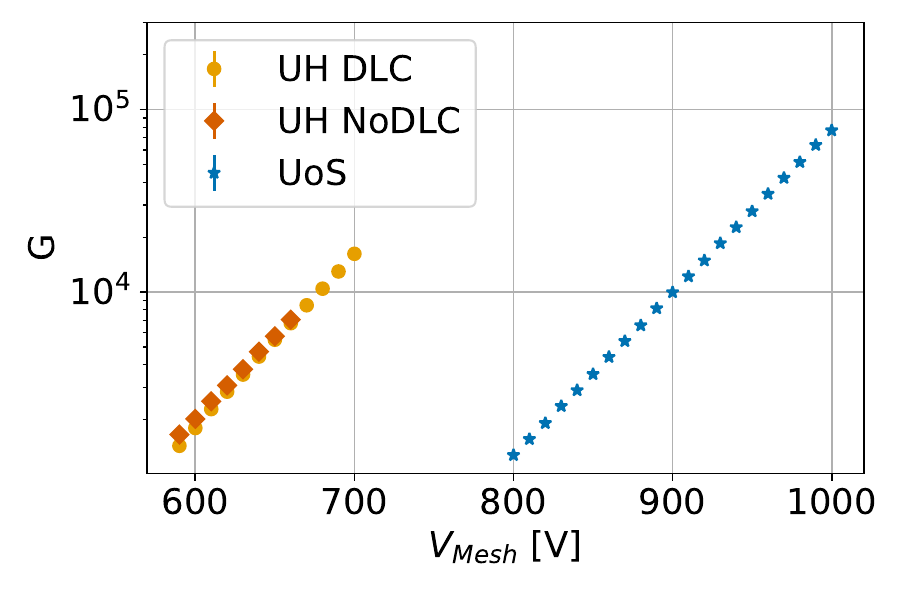}  
        \caption{ }
        \label{Vmesh}
    \end{subfigure}
    \begin{subfigure}{.495\textwidth}
        \centering
        % include second image
        \includegraphics[width=\linewidth,trim={.3cm 0.4cm 0.35cm 0.35cm},clip]{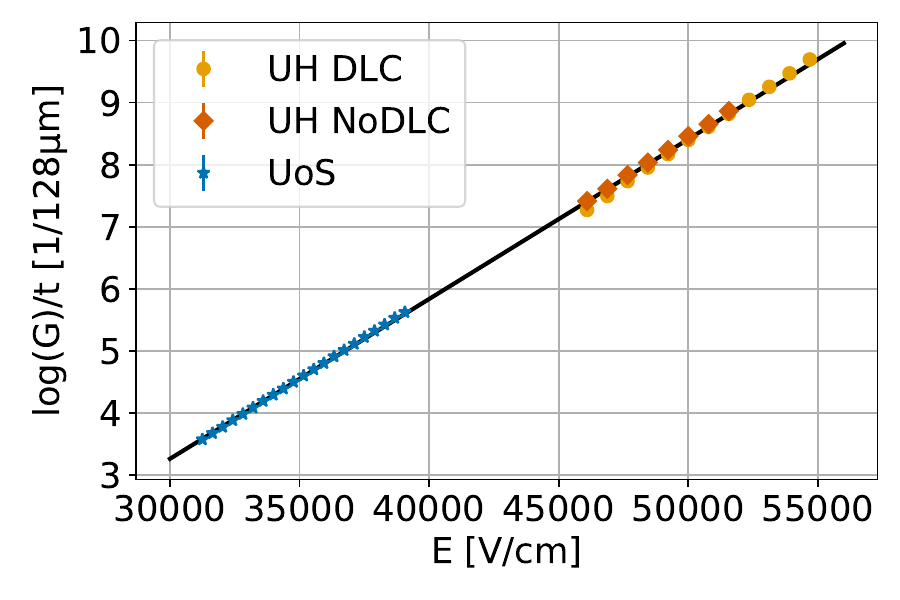}  
        \caption{ }
        \label{Efield}
    \end{subfigure}
    \caption{(a) Avalanche gain, as measured on the Micromegas by the PHA setup, versus Micromegas mesh voltage. The measurements are made with a Fe-55 source placed above the center of the readout plane for each detector. The error bars are smaller than the markers. (b) The same data are plotted to show the natural logarithm of the avalanche gain divided by the amplification gap thickness in units of 128\,$\upmu$m, versus the amplification field strength. The data are simultaneously fit to Equation~\ref{lin_fit}, indicated by the black line.}
\end{figure}

For each readout, we utilize the PHA setup to measure $G$ versus $V_\textrm{Mesh}$. We state positive values of $V_\textrm{Mesh}$ throughout this article, but it is understood that the actual values set on the HV supply are negative. As an example, Figure~\ref{PHAdist} displays the pulse-height distribution for the UH DLC detector operating at $V_\textrm{Mesh} = 630$\,V. The results are presented in Figure~\ref{Vmesh}. The error bars include statistical uncertainties and a 2.2\% systematic uncertainty, discussed in Section~\ref{PHA}. Since the UoS detector has a larger amplification gap, it requires higher $V_\textrm{Mesh}$ values and reaches larger gains before sparking. Sparking in this detector was observed beyond $V_\textrm{Mesh} = 1000$\,V, hence the largest observed $G$ is $76.8\times10^3$. The DLC layer enhances the sparking resistance of the UH DLC detector compared to the UH NoDLC. The UH DLC detectors begins sparking beyond $V_\textrm{Mesh} = 700$\,V, whereas the UH NoDLC detector begins beyond $V_\textrm{Mesh} = 660$\,V. Hence, the largest observed $G$ for the UH DLC and UH NoDLC detectors is $16.2\times10^3$ and $7.06 \times10^3$, respectively. Another advantage of the DLC layer is its ability to protect front-end chips from damage caused by sparking. Without a DLC layer, sparks often result in permanently damaged VMM channels. However, with a DLC layer, there was no permanent damage noted.  Apart from these differences, the relationship between $G$ and $V_\textrm{Mesh}$ is similar for the UH detectors.

To quantify the relationship between $G$ and $V_\textrm{Mesh}$, the data for all detectors are simultaneously fitted to
\begin{equation}
    \label{lin_fit}
    \log{(G)} = a t \left( E - E_o \right),
\end{equation}
where $a$ [1/V] and $E_o$ [V/cm] are fit parameters, $t$ [cm] is the amplification gap thickness, and $E = V_\textrm{Mesh} / t$ is the amplification field strength. The fit values are $a=(2.01 \times 10^{-2} \pm 2.70 \times 10^{-5}$)\,1/V and $E_o = (1.73 \times 10^{4} \pm 29.0$)\,V/cm. The fitted function is plotted alongside the data in Figure~\ref{Efield}. Here, the results are plotted in $\frac{\log{G}}{t}$ versus $E$ which illustrates how the data from all detectors are modelled by Equation~\ref{lin_fit}.

\begin{figure}[ht]
\begin{center}
\includegraphics[width=0.75\textwidth,trim={.35cm 0.4cm 0.35cm 0.3cm},clip]{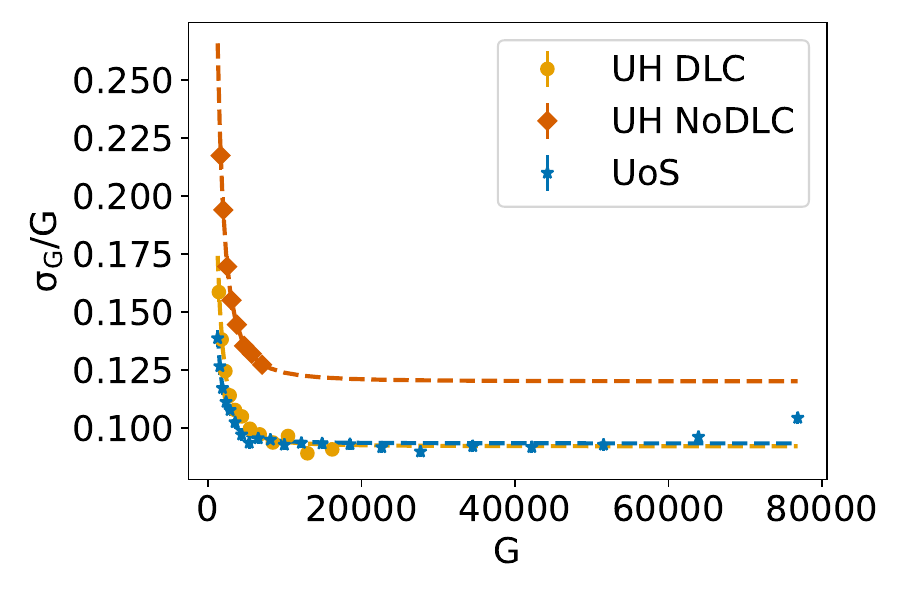}
%\framebox[4.0in]{$\;$}
%\fbox{\rule[-.5cm]{0cm}{4cm} \rule[-.5cm]{4cm}{0cm}}
\end{center}
\caption{The fractional avalanche gain resolution, $\sigma_G / G$, versus avalanche gain for all detectors. The data for each detector is fitted to Equation~\ref{invs} which is depicted by the dashed lines. The error bars are smaller than the markers.}
\label{gain_res}
\end{figure}

Another important detector performance metric is the fractional gain resolution, $\sigma_G/G$. To compare the detectors, $\sigma_G/G$ is plotted versus $G$ in Figure~\ref{gain_res}. The results for each detector is fit to
\begin{equation}
    \sigma_G/G = \sqrt{ \left( \frac{\beta}{G} \right)^2 + \gamma^2 }.
\label{invs}
\end{equation}
Above, $\beta$ and $\gamma$ are dimensionless fit parameters. The fit value of the asymptotic fractional gain resolution, $\gamma$, is $0.092 \pm 0.001(\textrm{stat}) \pm 0.008(\textrm{syst})$, $0.120 \pm 0.002(\textrm{stat}) \pm 0.006(\textrm{syst})$, and $0.093 \pm 0.001 (\textrm{stat}) \pm 0.004 (\textrm{syst})$ for the UH DLC, UH NoDLC, and UoS detectors, respectively. The systematic uncertainties were evaluated by repeating the PHA fits without the 2nd order polynomial background function. The value of $\gamma$ is notably higher for UH NoDLC, indicating that the omitting the DLC layer adversely impacts $\sigma_G/G$. 

To assess the positional dependence of the detectors, $G$ is measured with the uncollimated Fe-55 source placed at each corner of the sensitive area of the detectors. Doing so, the X-ray conversion events are well-contained above a single quadrant of the readouts. A Micromegas mesh voltage of $V_\textrm{Mesh} = 660$\,V is applied for the UH detectors and $V_\textrm{Mesh} = 1000$\,V for the UoS detector. The results are summarized in Table~\ref{quadrants}, where the quadrants, labeled $a$, $b$, $c$, and $d$, are defined in Figure~\ref{top_view}. The positional variation in $G$ is 4\%, 5\%, and 6\% for the UoS, UH DLC, and UH NoDLC detectors, respectively.

\begin{table}[h!]
\centering
\begin{tabular}{|c| c| c |c |c|c|c|} 
 \hline
 Detector & $G_\textrm{quad.~a}$ & $G_\textrm{quad.~b}$ & $G_\textrm{quad.~c}$ & $G_\textrm{quad.~d}$ & $V_\textrm{Mesh}$ [V] & $ {\sigma / \mu}$ \\ [0.5ex] 
 \hline
 UH DLC & $6663$ & $6118$ & $6843$ & $6617$ & $660$ & $0.05$ \\ 
 UH NoDLC & $6664$ & $7085$ & $6338$ & $7258$ & $660$ & $0.06$ \\ 
 UoS & $76664$ & $77865$ & $83716$ & $82615$ & $1000$ & $0.04$ \\  [1ex] 
 \hline
\end{tabular}
\caption{Micromegas avalanche gain, measured with the PHA setup and an Fe-55 source for each quadrant of the detectors. The two rightmost columns displays the Micromegas mesh voltage used, and the fractional gain variation over the four quadrants.}
\label{quadrants}
\end{table}

\section{Characterization via Strip Readout}
\label{VMM}

To obtain spatial information, the strips are read out using the VMM hybrids and SRS DAQ system discussed in Section~\ref{exp-setup}. The UH detectors require four VMM hybrids on each axis, as shown in Figure~\ref{MM_pic}. The UoS detector requires only three VMM hybrids per axis as it has a wider strip pitch and hence fewer strips. The same VMM hybrids are used to readout all the detectors and they are placed in the same positions (where possible) in order compare the readouts with identical electronics.

The VMM/SRS DAQ system~\cite{Scharenberg_2022,PFEIFFER2022166548,LUPBERGER201891} is controlled by the \texttt{VMM Slow Control} Software (\texttt{VMMSC})~\cite{vmmsc}. Each VMM channel combines an analogue and a digital section. The analogue part consists of a charge sensitive preamplifier followed by a shaper, discriminator, and peak finder. The electronic gain of the preamplifier is set to 4.5\,mV/fC (9\,mV/fC) on channels connected to upper (lower) strips because the induced signal is smaller on the lower strips versus the upper strips. The peaking time of the shaper is set to 200\,ns. The discriminator is used to operate the system in a self-triggered continuous mode where a signal is digitally processed if it surpasses a set threshold. In this case, the peak amplitude that is identified by the peak finder is transferred to a 10-bit ADC. The time of the peak is found with respect to a 40\,MHz clock referred to as the Bunch Crossing Clock (CKBC), hence this output is called the BCID. A fine time correction is captured by a voltage ramp that starts at the time of the peak and stops at the falling edge of the next CKBC signal. The slope of this Time-to-Amplitude Converter (TAC) is set to 60\,ns and it produces an 8-bit time detector output (TDO). The slope setting and TDO value give a fine timestamp correction to the BCID. The time required for a channel to digitize a hit is 250\,ns; hence, the maximum rate is 4\,Mhits/s. The UoS readout exhibits a higher rate of noise hits when the threshold is the same as for the other readouts. To mitigate this, we use a lower electronic gain setting of 1.0\,mV/fC across all VMMs for the UoS readout. Aside from this, all other settings are consistent throughout the detectors.

\subsection{Calibration}
\label{calib}

Each VMM contains two global 10-bit DACs, meaning that they affect all 64-channels: The ``pulser DAC'' sets the test voltage used to generate internal test pulses, and the second ``threshold DAC'' drives the threshold setting of all channels. We first perform the threshold and pulser DAC calibration scans on \texttt{VMMSC} to characterize the voltage versus DAC setting, as discussed in Ref.~\cite{Scharenberg_2022}. Next, the pedestal voltage of each channel is measured and the mean pedestal is computed for each VMM. Using the mean pedestal value and the linear relationship between threshold DAC and the threshold voltage, we find the DAC setting corresponding to 100\,mV above pedestal for each VMM. This value is used to set the threshold DAC of each VMM.

\begin{figure}[ht]
    \begin{subfigure}{.495\textwidth} 
        \centering
        % include first image
        \includegraphics[width=\linewidth,trim={.3cm 0.4cm 0.35cm 0.35cm},clip]{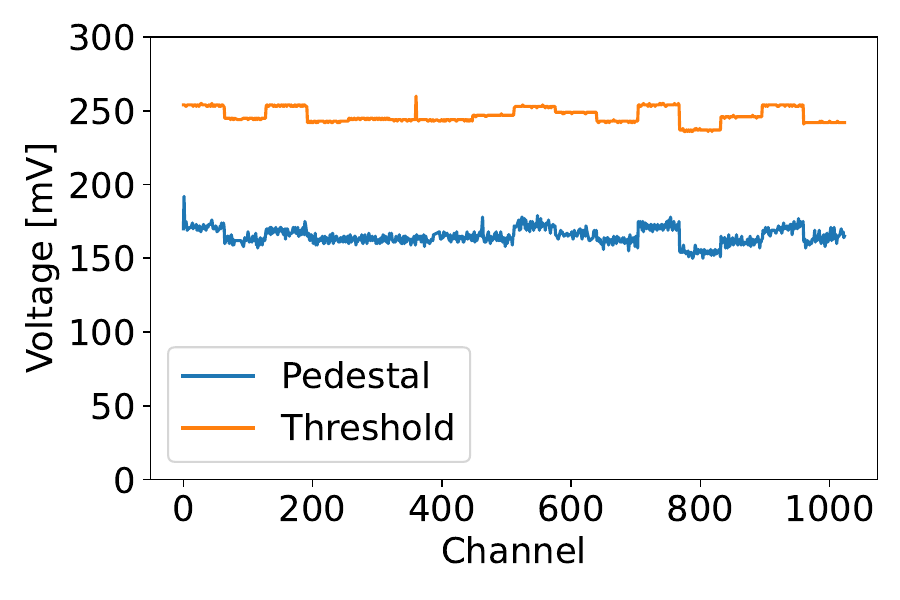}  
        \caption{ }
        \label{calibs}
    \end{subfigure}
    \begin{subfigure}{.495\textwidth}
        \centering
        % include second image
        \includegraphics[width=\linewidth,trim={.3cm 0.4cm 0.35cm 0.35cm},clip]{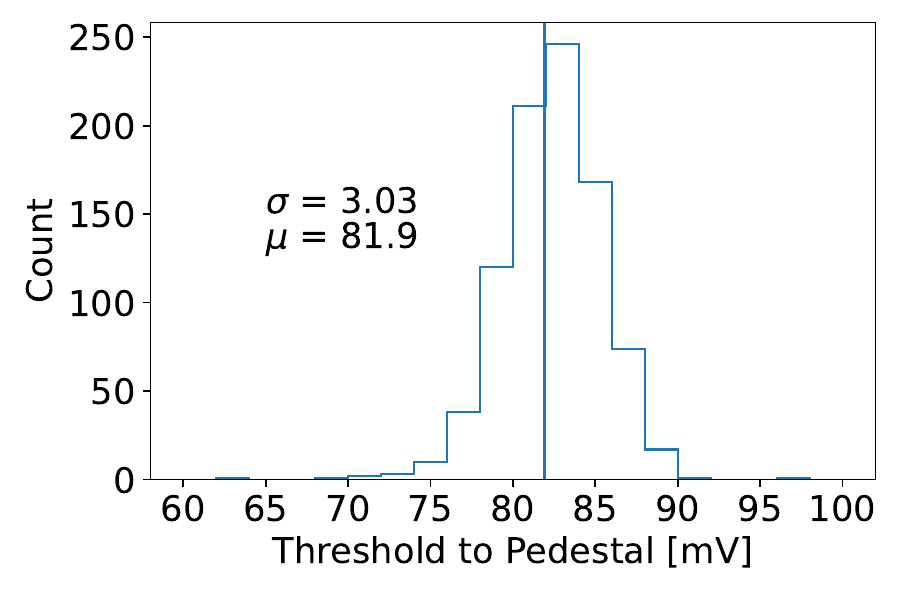}  
        \caption{ }
        \label{calibs2}
    \end{subfigure}
    \caption{Result of channel-level threshold calibration of the eight VMM front end readout ASCICs. (a) Pedestal and threshold versus channel number after the threshold DAC calibration. (b) Histogram of the distance from the threshold to the pedestal, in mV, for every channel after the threshold DAC calibration.}
\end{figure}

Although a global threshold is set for each VMM, each ASIC also exhibits variations across its 64 channels. To compensate for this, each VMM is equipped with 64 local 5-bit threshold trimming DACs. The threshold DAC calibration (provided by \texttt{VMMSC}) is used to fine-tune the 5-bit trimmer values to minimize threshold dispersion. Post-calibration, the final threshold voltage for each channel is measured. The threshold and pedestal for all channels is illustrated in Figure~\ref{calibs}. The difference, in mV, between the threshold and pedestal for each channel is also depicted in Figure~\ref{calibs2}. The calibrated thresholds averaged 81.9\,mV above the pedestal after the threshold DAC calibration.

The final step in calibration involves utilizing the ADC and BCID/TDC calibrations available in \texttt{VMMSC} to calibrate the ADC and timing response of every channel using internal calibration pulses, as described in Ref.~\cite{Scharenberg_2022}.

\subsection{Noise}
\label{Noise}

The \texttt{VMMSC}~\cite{vmmsc} software has a built-in method for measuring the noise~\cite{kolanoski2020particle}. VMM channels are pulsed repeatedly for a range of threshold settings, and the resulting hit rate versus threshold setting is recorded. The resulting ``S-curve'' is fit to the complementary error function to determine the noise, $\sigma_\textrm{noise}$. The noise, initially measured in mV, can be divided by the electronic (VMM) gain to determine the $\sigma_\textrm{noise}$ in electrons.

\begin{figure}[ht]
    \begin{subfigure}{.495\textwidth} 
        \centering
        % include first image
        \includegraphics[width=\linewidth,trim={0.2cm 0cm 0.2cm 0.4cm},clip]{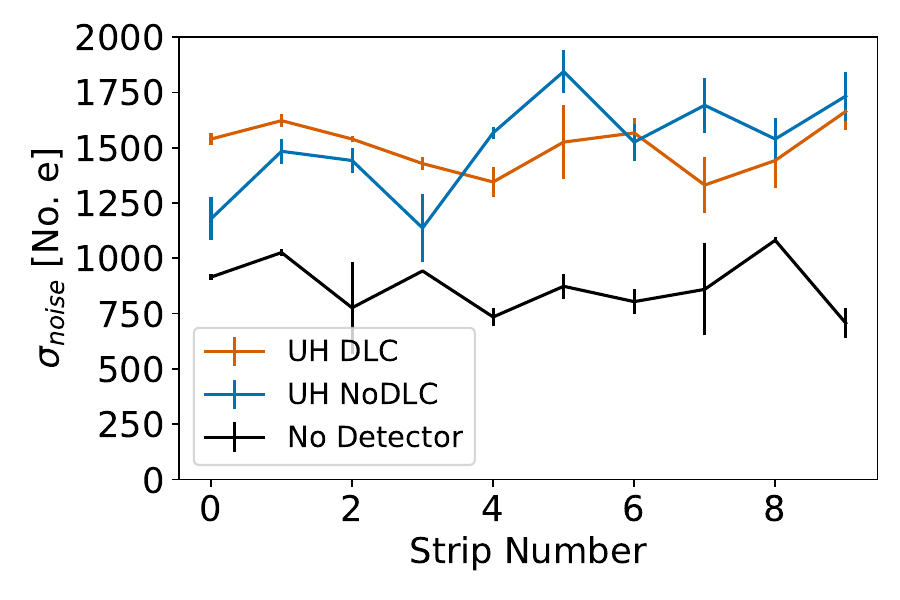}  
        \caption{}
        \label{S_curve_up}
    \end{subfigure}
    \begin{subfigure}{.495\textwidth}
        \centering
        % include second image
        \includegraphics[width=\linewidth,trim={0.2cm 0cm 0.2cm 0.4cm},clip]{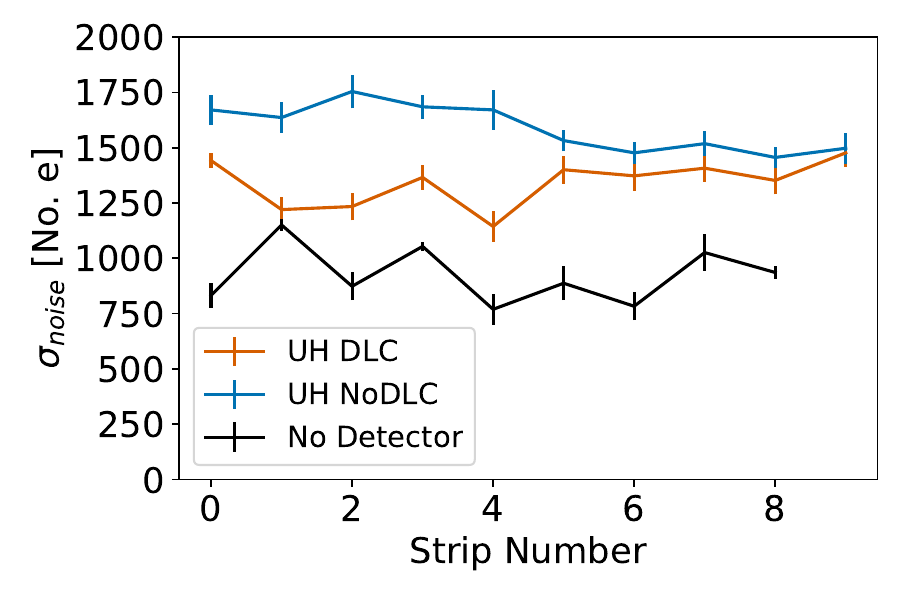}  
        \caption{}
        \label{S_curve_low}
    \end{subfigure}
    \caption{VMM noise level, $\sigma_\textrm{noise}$, when connected to the UH DLC and UH NoDLC detector strips. The noise is measured via threshold scans and S-curve fits for 10 strips, connected to two VMMs. Error bars correspond to the statistical fitting uncertainty of S-curves. For comparison, the measurement is repeated with the VMM hybrids not connected to any detector. (a) Upper strips. (b) Lower strips.}
    \label{S_curve}
\end{figure}

The S-curve implementation in \texttt{VMMSC} has known instabilities~\cite{Scharenberg_2022}. To circumvent these, we measure the noise for only four VMMs, two in $x$ and two in $y$. For each VMM, an S-curve is measured on five channels and all other channels are masked. The results for the UH DLC and UH NoDLC readouts are presented in Figure~\ref{S_curve}. With the detectors disconnected from the VMMs, we find $\sigma_\textrm{noise} = 897 \pm 122$\,electrons for all channels. This is consistent with independent VMM noise characterization measurements at CERN~\cite{noise}. With the detectors connected, we find $\sigma_\textrm{noise} = 1487 \pm 165$\,electrons for both the lower and upper strips of both readouts. This noise level is expected for a capacitance of $\approx 50$\,pF connected to the VMM~\cite{noise}. We expect the noise level to vary with the strip capacitance, which varies with strip dimensions. While some such variation is observed, the statistical significance is low. The most significant observation is that for the lower strips, the observed noise is 6\,$\sigma$ lower in detectors with a DLC layer than in detectors without a DLC layer. In future work, we plan a detailed follow-up investigation, once we have a more stable noise scan and an improved grounding scheme implemented.

\subsection{Gain, Gain Resolution, and Charge Sharing}
\label{VMM_Fe55}

\begin{figure}[ht]
\begin{center}
\includegraphics[width=0.95\textwidth,trim={.4cm .3cm 0.4cm 1cm},clip]{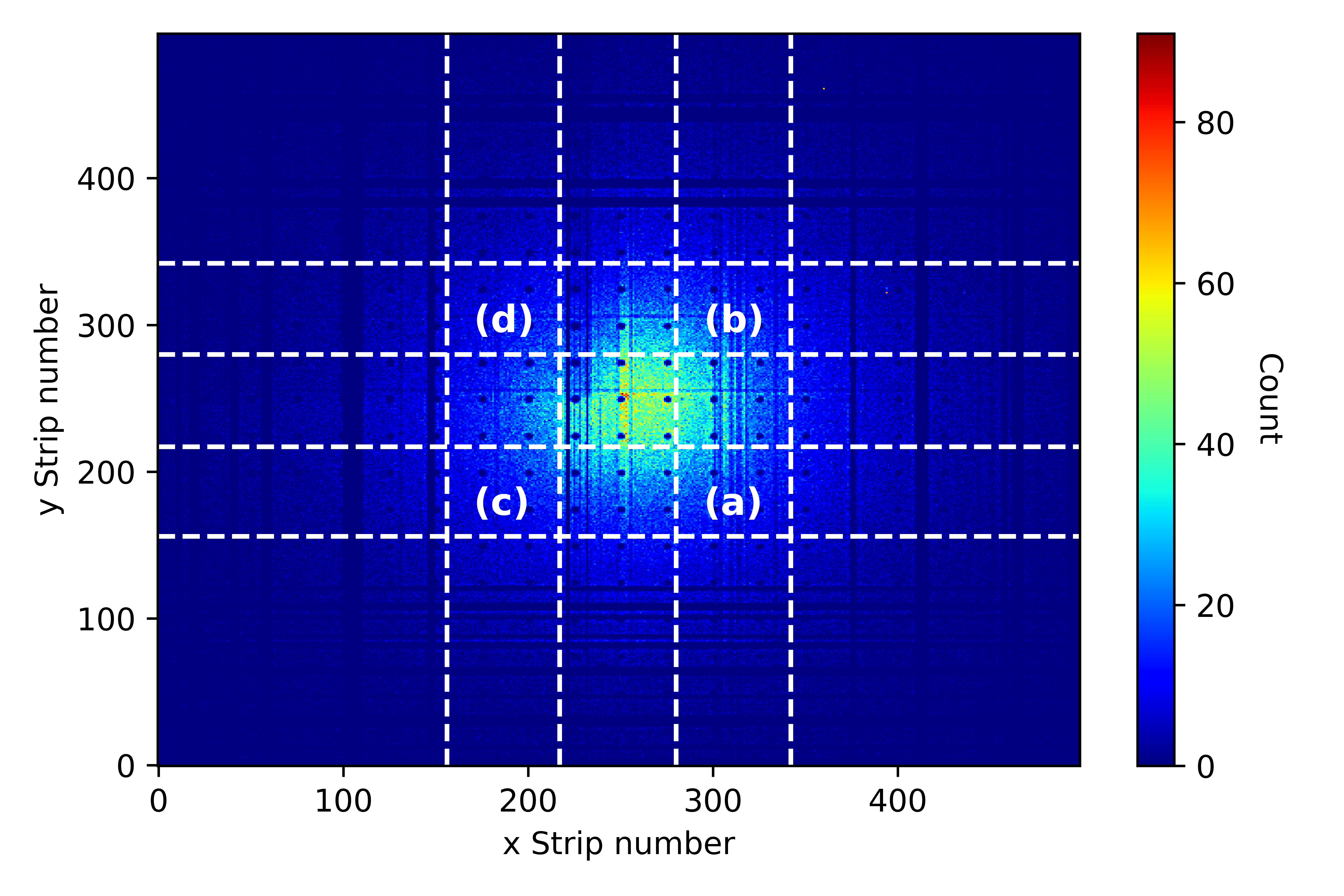}
%\framebox[4.0in]{$\;$}
%\fbox{\rule[-.5cm]{0cm}{4cm} \rule[-.5cm]{4cm}{0cm}}
\end{center}
\caption{Two-dimensional histogram displaying the center of charge positions of Fe-55 X-ray conversion events, as detected by the UH DLC detector with $V_\textrm{mesh} = 700$ V. Dashed lines indicate the boundaries of individual VMM front-end chips used for analysis. The analysis of digitized strip data considers only events fully contained in either region $a$, $b$, $c$, or $d$.  Each of these regions has a unique upper strip width. The empty circles are a shadow image of the pillars used to support the Micromegas mesh.}
\label{fe55_clusters}
\end{figure}

VMM strip data is collected for all detectors with the radioactive Fe-55 source placed above the center of the readout plane, as shown in Figure~\ref{MM_Fe55}. Data collection is carried out in 5-minute intervals for each value of $V_\textrm{mesh}$. The initial data processing involves clustering raw hits into events using the \texttt{VMM-sdat} software~\cite{vmm-sdat}. This step creates events by matching raw $x$ and $y$ hits based on the parameters summarized in the Fe-55 column of Table~\ref{vmmsdat_params1}. To illustrate how the Fe-55 events illuminate the readout, the distribution of Fe-55 events on the UH DLC readout with $V_\textrm{mesh} = 700$\,V is illustrated in Figure~\ref{fe55_clusters}. This analysis is confined to Fe-55 clusters with hits that are entirely contained in four specific regions, labeled as $a$, $b$, $c$, and $d$ in Figure~\ref{fe55_clusters}. The regions are subsets of the identically labeled quadrants of the readout (see Figure~\ref{top_view}), chosen such that they are read out by a single VMM in both the $x$ and $y$ dimensions. For the UH detectors, each of the quadrants have a different upper strip width, see Table~\ref{MM_config}.

\begin{table}[h!]
\centering
\begin{tabular}{|c| c|c|c|} 
 \hline
 Param. & Description & Fe-55  & Po-210  \\ [0.5ex] 
 \hline
 cs & Min. cluster size per plane & $2$ & $3$ \\ 
 ccs & Min. cluster size in both planes & $4$ & $6$ \\ 
 mst & Max. no. missing strips in strip sorted vector & $1$ & $15$ \\ 
 dt & Max. time b/w strips in time sorted vector & $200$ & $200$ \\ 
 spc & Max. time span of cluster in 1D & $1500$ & $1500$ \\ 
 dp & Max. time b/w matched clusters in $x$ \& $y$ & $200$ & $400$ \\ 
 crl & Lower limit on charge in plane 0 / plane 1 & 0 & 0 \\ 
 cru  & Upper limit on charge in plane 0 /  plane 1 & 1000 & 1000  \\  [1ex] 
 \hline
\end{tabular}
\caption{The \texttt{VMM-sdat}~\cite{vmm-sdat} parameters used to cluster photoelectric events induced by a radioactive Fe-55 source and alpha events emitted by a radioactive Po-210 source. The default values are used for all unspecified parameters.}
\label{vmmsdat_params1}
\end{table}

\begin{figure}[!h]
    \centering   %
\setkeys{Gin}{width=\linewidth}
\begin{subfigure}{.49\textwidth}
        \includegraphics[width=\linewidth,trim={.4cm .5cm 0.3cm 0.45cm},clip]{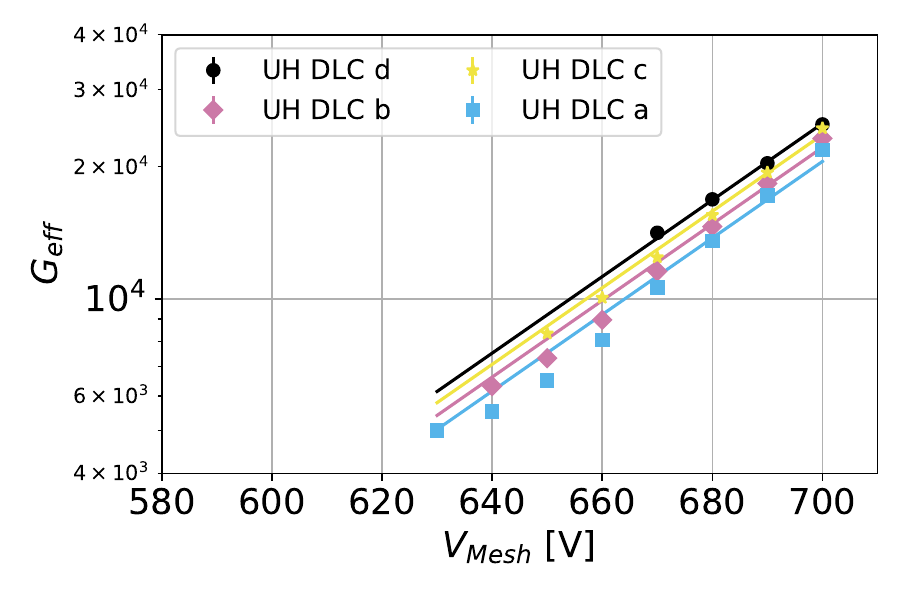}    \caption{}
        \label{VMM_Fe55_UHDLC}
\end{subfigure}
\hfil
\begin{subfigure}{.49\textwidth}
        \includegraphics[width=\linewidth,trim={.4cm .5cm 0.3cm 0.45cm},clip]{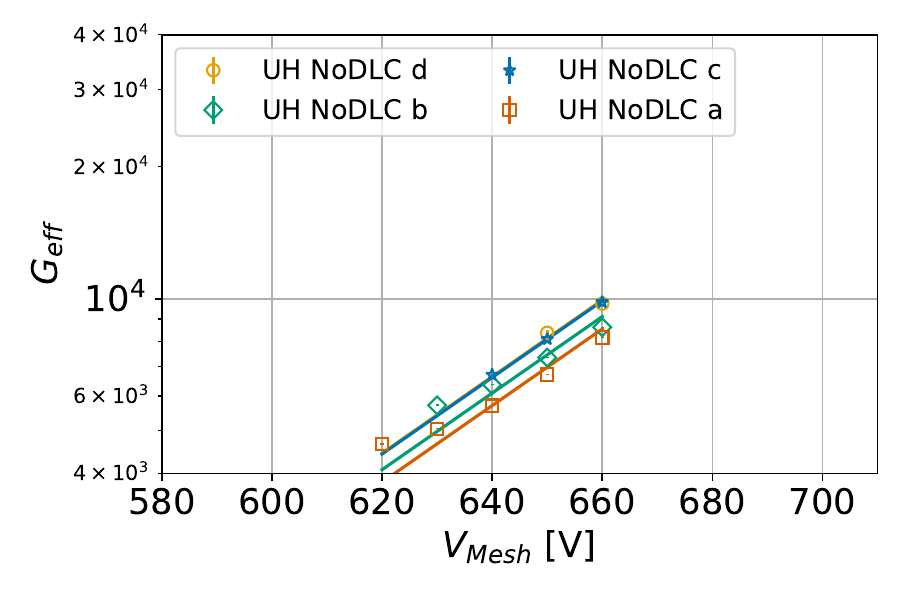}  
  \caption{}
        \label{VMM_Fe55_UHNoDLC}
\end{subfigure}

\begin{subfigure}{.49\textwidth}
        \includegraphics[width=\linewidth,trim={.4cm .5cm 0.3cm 0.35cm},clip]{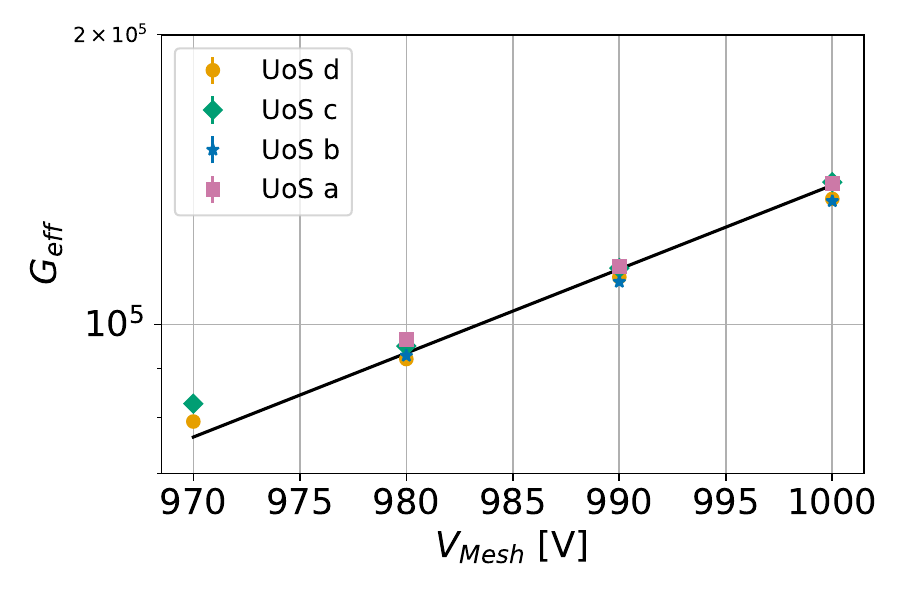}  
  \caption{}
    \label{VMM_Fe55_UoS}
\end{subfigure}%
\caption{Effective gain measured on the strips using VMM/SRS versus Micromegas mesh voltage. (a) For all quadrants of the UH DLC detector. (b) For all quadrants of the UH NoDLC detector. (c) For all quadrants of the UoS detector. Statistical error bars are smaller than the markers in all plots. The solid lines illustrate the fit of Equation~\ref{M_fit} to the data.}
\label{comparison-blood-vessels}
\end{figure}

The VMM/SRS DAQ system provides an ADC scale which is linearly proportional to the detected charge after avalanche amplification. The appropriate method for measuring avalanche gain is discussed in Section~\ref{PHA_results}; here, we estimate the effective gain at the strip level by using the conversion 1\,ADC $\approx$ 1\,mV. Next, we divide by the electronic gain setting of the relevant VMM channel to obtain a measurement in units of charge. The charge is summed over all hits in a clustered event then divided by $\textrm{N}_{exp}$ to obtain a measurement of the effective gain for a single Fe-55 X-ray conversion event. Following the same approach as Section~\ref{PHA}, $G_\textrm{eff}$ and $\sigma_{G_\textrm{eff}}$ are obtained by fitting Equation~\ref{crystalball} to the gain distributions. Here, $G_\textrm{eff}$ denotes the effective gain measured on the strips using VMM/SRS. Plots of $G_\textrm{eff}$ versus $V_\textrm{mesh}$ for all quadrants in all detectors are presented in Figures~\ref{VMM_Fe55_UHDLC}--~\ref{VMM_Fe55_UoS}.  The approximation 1\,ADC $\approx$ 1\,mV appears in the y-axis of these figures; however, all conclusions drawn from this data are stated as ratios that are independent of this conversion. Note that configurations with wider $y$ strips and hence a weaker signal on the $x$ strips (e.g. quadrant d) require larger $V_\textrm{Mesh}$ values to observe hits on both $x$ and $y$ strips so that $G_\textrm{eff}$ can be measured. 

Although our method for estimating gain with the VMM/SRS system is a rough approximation, since the ADC scale is linearly proportional to charge, our results should match those presented in Section~\ref{PHA}, up to a multiplicative factor (M). To determine M, the data for each quadrant is fit to 
\begin{equation}
    \label{M_fit}
    G_\textrm{eff} = M \exp{ \left[ a t \left( E - E_o \right) \right] },
\end{equation}
which is simply the exponentiation of Equation~\ref{lin_fit} multiplied by a dimensionless fit parameter M. Above, $a$ and $t$ are the fit values obtained in Section~\ref{PHA_results}. The fits are depicted by the solid lines in Figures~\ref{VMM_Fe55_UHDLC}--~\ref{VMM_Fe55_UoS}. At low $V_\textrm{Mesh}$, the data points noticeably deviate from the fits. This is because at low $V_\textrm{Mesh}$ hits fall below threshold more frequently. In this regime, the data is not well modeled by Equation~\ref{M_fit}.

For the UH DLC detector, we find that M = 1.37, 1.47, 1.57, and 1.67 for quadrants $a$, $b$, $c$, and $d$, respectively. The variation in M over different quadrants reveals how the effective gain is influenced by the upper strip width. For example, quadrant $d$, which has an upper strip width of 100\,$\upmu$m, observes an effective gain that is 21.9\% larger than quadrant $a$, which has an upper strip width of 40\,$\upmu$m. For the UH NoDLC detector, we find that M = 1.27, 1.35, 1.47, and 1.47 for quadrants $a$, $b$, $c$, and $d$, respectively. In this case quadrant $d$ observes an effective gain that is 15.7\% larger than quadrant $a$. For the UoS detector which only has one strip configuration, the mean over all four quadrants is fit to obtain M = 1.91.

\begin{figure}[ht]
\begin{center}
\includegraphics[width=0.99\textwidth,trim={0.4cm 0.5cm 0.35cm 0.4cm},clip]{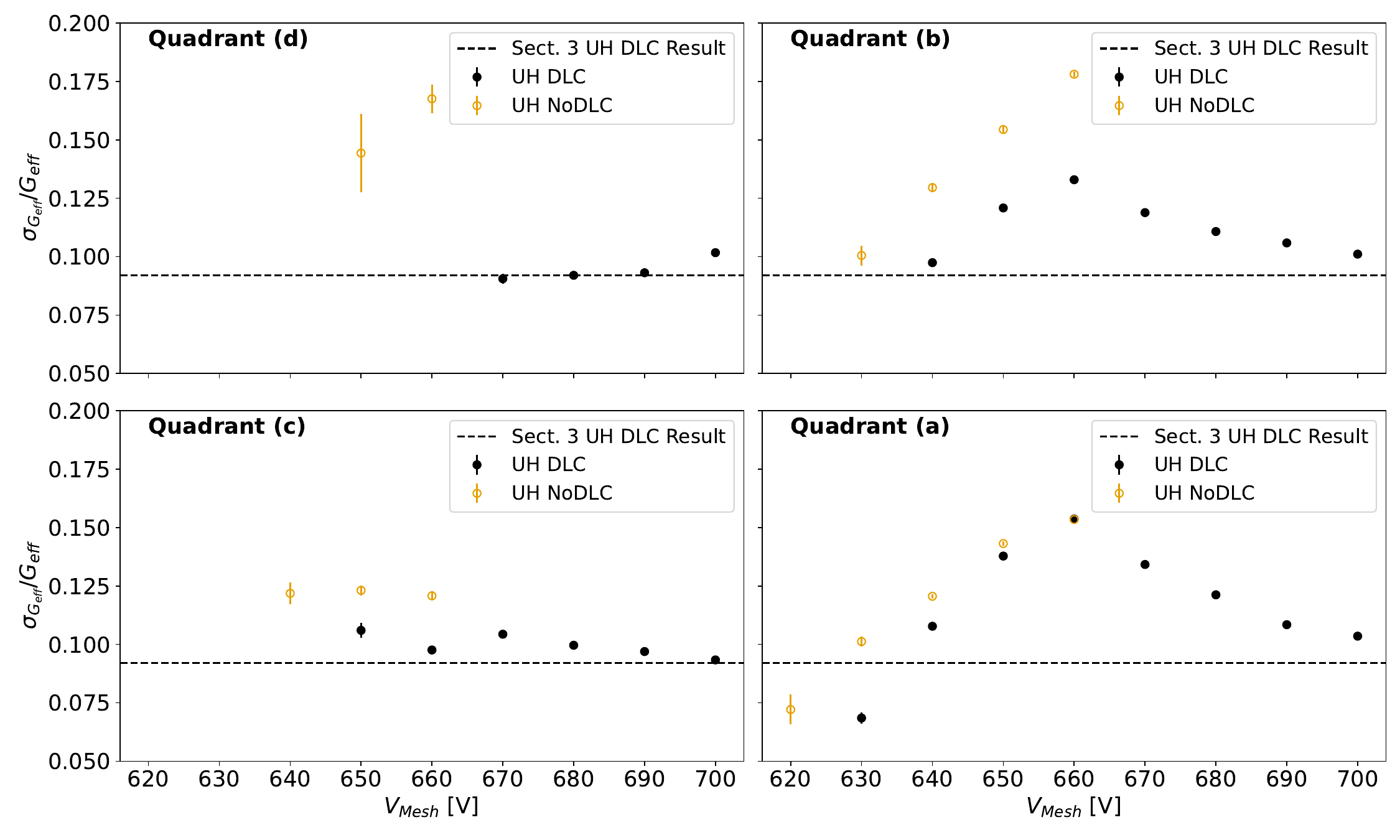}
%\framebox[4.0in]{$\;$}
%\fbox{\rule[-.5cm]{0cm}{4cm} \rule[-.5cm]{4cm}{0cm}}
\end{center}
\caption{The fractional effective gain resolution as measured on the strips using VMM/SRS versus Micromegas mesh voltage for all quadrants of the UH detectors. A label is included in each subplot indicating the quadrant that it corresponds to. The dashed black line indicates the asymptotic behaviour found using the PHA setup to measure the avalanche gain on the Micromegas mesh.}
\label{VMM_GainRes_UH}
\end{figure}

\begin{figure}[ht]
\begin{center}
\includegraphics[width=0.7\textwidth,trim={.4cm .4cm 0.3cm 0.4cm},clip]{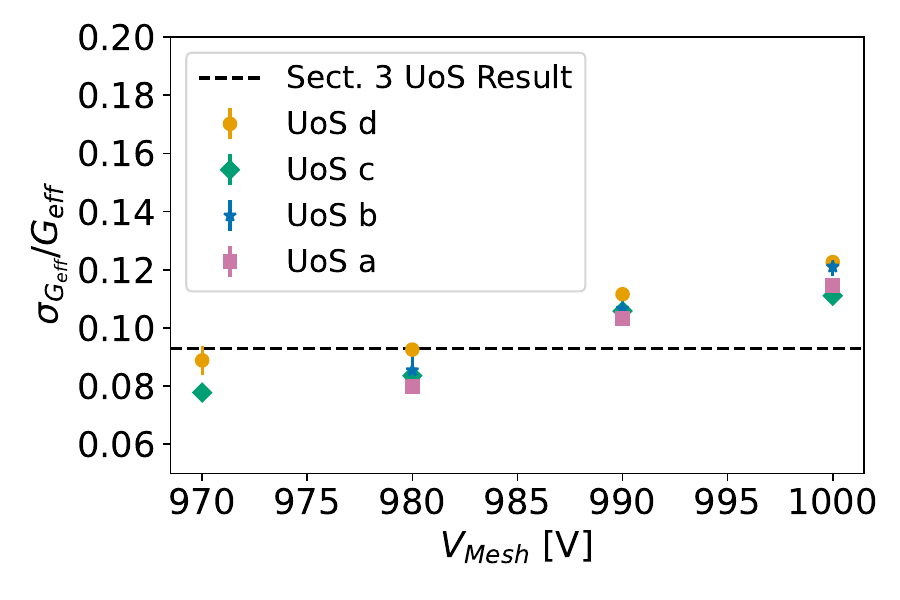}
%\framebox[4.0in]{$\;$}
%\fbox{\rule[-.5cm]{0cm}{4cm} \rule[-.5cm]{4cm}{0cm}}
\end{center}
\caption{The fractional effective gain resolution as measured on the strips using VMM/SRS versus Micromegas mesh voltage for all quadrants of the UoS detector. The dashed black line indicates the asymptotic fractional avalanche gain resolution reported in Section~\ref{PHA} for the UoS detector.}
\label{VMM_GainRes_UoS}
\end{figure}

Figures~\ref{VMM_GainRes_UH} and~\ref{VMM_GainRes_UoS} show $\sigma_{G_\textrm{eff}} / G_\textrm{eff}$ versus $V_\textrm{mesh}$, across all quadrants in every detector. The motivation for these measurements is to compare the asymptotic behavior of this effective gain resolution at high gain against the asymptotic behavior of the avalanche gain resolution (Figure~\ref{gain_res}). If the former is larger, this would indicate that the detection of charge in the strips deteriorates the charge measurement. 
There is some distracting behavior observed in Figure~\ref{VMM_GainRes_UH} at low gain (but this does not affect the main finding): for example, in quadrant $a$ of the UH DLC detector, $\sigma_{G_\textrm{eff}} / G_\textrm{eff}$ increases with $V_\textrm{Mesh}$ up to $V_\textrm{Mesh} = 660$\,V, beyond which it begins to plateau to the asymptotic fractional gain resolution measured in Section~\ref{PHA}. This pattern is attributable to the analysis being performed on digitized data. At lower $V_\textrm{Mesh}$ values, hits fall below the detection threshold, leading to an artificial reduction in $\sigma_{G_\textrm{eff}}$. As $V_\textrm{Mesh}$ increases, the fraction of hits below threshold decreases, resulting in the expected plateauing behavior. The asymptotic fractional gain resolution measured in Section~\ref{PHA} for the UH DLC detector is depicted by the black dashed line in Figure~\ref{VMM_GainRes_UH}. For the UH NoDLC detector, sparking occurs before the $\sigma_{G_\textrm{eff}} / G_\textrm{eff}$ ratio begins to plateau. For the UoS detector (Figure~\ref{VMM_GainRes_UoS}), a consistent $\sigma_{G_\textrm{eff}} / G_\textrm{eff}$ ratio is observed across all quadrants as the strip configuration is uniform. We do not observe the expected plateauing behavior in the detector; this is likely due to the low electronic gain, resulting in hits falling below the detection threshold across all $V_\textrm{Mesh}$ values explored. The main finding is that the asymptotic effective gain resolution is very close to the asymptotic avalanche gain resolution. We do not see any evidence for any broadening of the gain resolution due to the strips. This implies that strip readout as tested here should not negatively affect energy resolution of a future detector.

\begin{figure}
 \centering
 \begin{subfigure}[h]{0.49\textwidth}
     \centering
     \includegraphics[width=\textwidth]{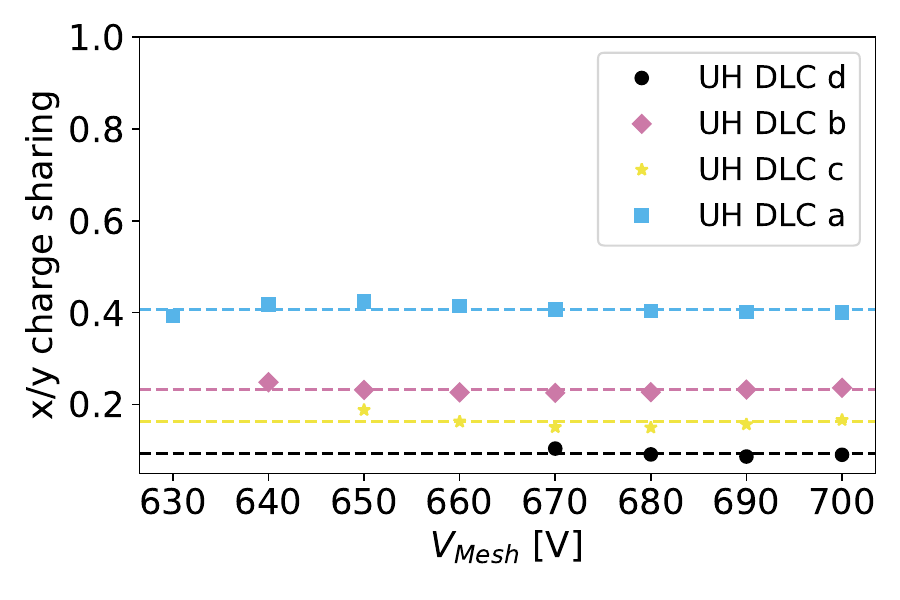}
     \caption{}
     \label{VMM_CS_UHDLC}
 \end{subfigure}
 \hfill
 \begin{subfigure}[h]{0.49\textwidth}
     \centering
     \includegraphics[width=\textwidth]{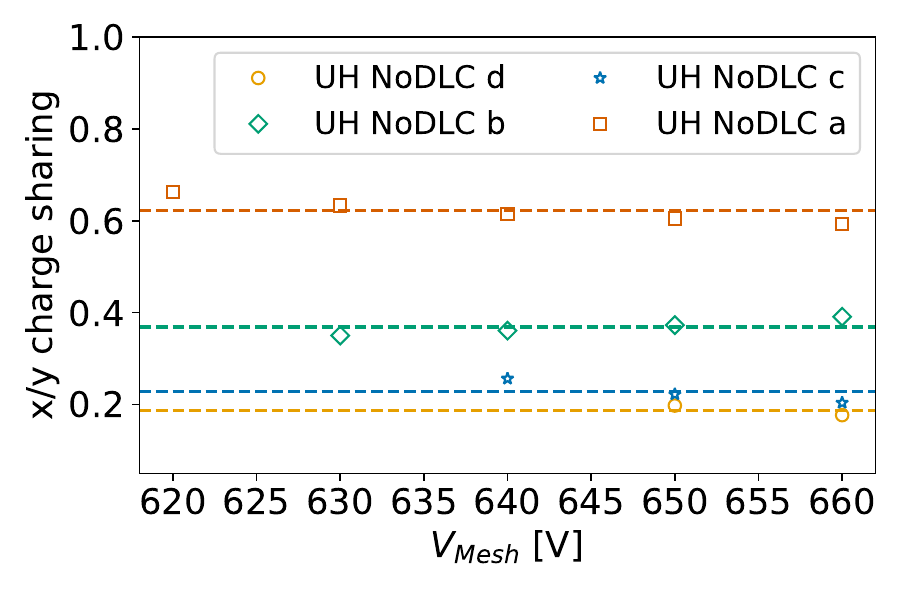}
     \caption{}
     \label{VMM_CS_UHNoDLC}
 \end{subfigure}
 \hfill
 \begin{subfigure}[h]{0.49\textwidth}
     \centering
     \includegraphics[width=\textwidth]{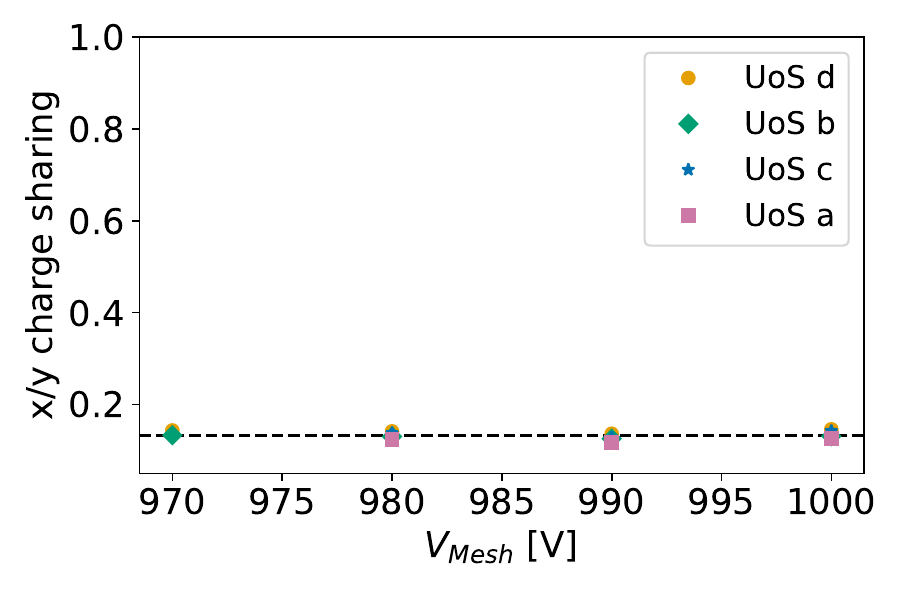}
     \caption{}
     \label{VMM_CS_UoS}
 \end{subfigure}
 \hfill
 \begin{subfigure}[h]{0.49\textwidth}
     \centering
     \includegraphics[width=\textwidth]{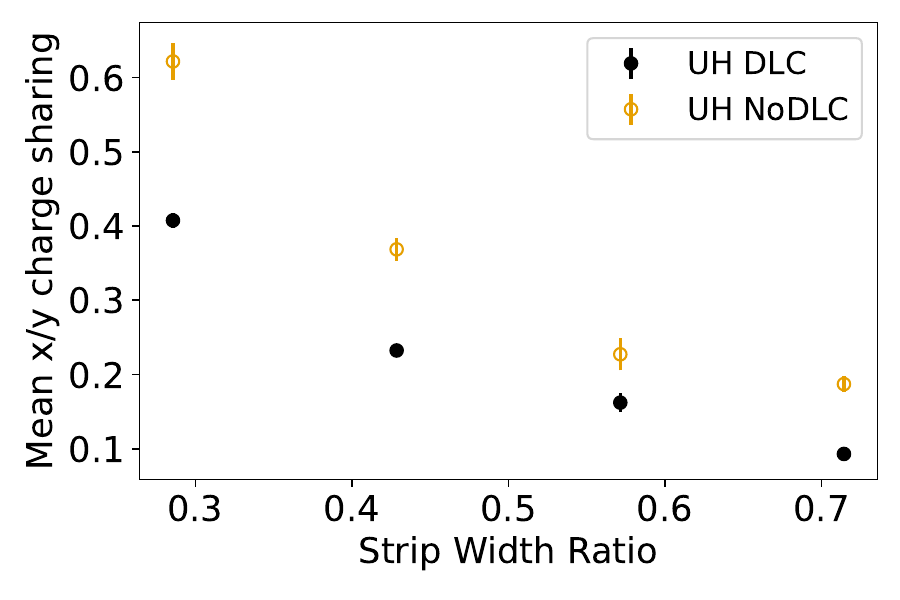}
     \caption{}
     \label{CS_trend}
 \end{subfigure}
    \caption{Ratio of the effective charge detected on the lower strips to that detected on the upper strips, denoted as ``$x/y$ charge sharing'', versus Micromegas mesh voltage for the UH DLC (subplot a), UH NoDLC (subplot b), and UoS (subplot c) detectors. The marker color indicates quadrant. For the UH DLC and NoDLC detectors, different quadrants have difference upper strip widths, detailed in Table~\ref{MM_config}. The UoS detector has the same strip widths across all quadrants. The mean charge sharing, after averaging over $V_\textrm{Mesh}$, is indicated with a dashed line for each quadrant of the UH detectors and all quadrants of the UoS detector. In subplot (d), the mean and standard deviation is plotted versus the strip width ratio corresponding to each quadrant of the UH detectors.}
    \label{VMM_CS}
\end{figure}

Another important quantity is the $x/y$ charge sharing (CS), which we define as the mean ratio of the number of electrons detected on the lower strips to the upper strips, over clustered Fe-55 events. This quantity is plotted versus $V_\textrm{Mesh}$ for the UH DLC, UH NoDLC, and UoS detectors in Figure~\ref{VMM_CS}. In all detectors, CS is a function of strip geometry and mostly independent of $V_\textrm{Mesh}$. For the UH detectors, it is evident that quadrants with thinner upper strips have CS values closer to one. Comparing UH DLC to UH NoDLC, it is notable that the inclusion of a DLC layer reduces CS. As expected, a uniform CS is measured across the quadrants of the UoS detector. Perfect charge sharing is indicated by $\textrm{CS} = 1$, this is ideal for 3D reconstruction which depends equally on information from the $x$ and $y$ axis. In Figure~\ref{CS_trend}, we display the mean CS versus the ratio of the upper strip to lower strip width for the UH detectors. The mean CS is 0.41, 0.23, 0.16, and 0.09 for UH DLC quadrant $a$, $b$, $c$, and $d$, respectively. For UH NoDLC the mean CS values are 0.62, 0.37, 0.23, and 0.19, in the same order. In both cases, the results suggest that thinner upper strips are required to reach $\textrm{CS}=1$. The lower strips are already at the maximum practical width, due to minimum spacing achievable with printed circuit board manufacturing techniques involving solid photo resist. Broader lower strips would require alternative manufacturing techniques, such as thin film photolithography with liquid resist, which would significantly increase costs~\cite{rui}.

\subsection{3D Reconstruction}
\label{3Drecon}
An ideal gas TPC would reconstruct the 3D position of each primary electron in an ionization event, and the time that the event took place. Knowledge of the full 3D primary charge topology maximizes the particle identification performance~\cite{Ghrear:2020pzk,Schueler:2022lvr,baracchini2020identification}. For recoil events, this topology can also be used to reconstruct the 3D (vector) direction of the recoil~\cite{Hedges:2021dgz,ghrear2024deep,10.1093/ptep/ptad120,Tao:2019wfh}, which in combination with the event time maximizes sensitivity of proposed experiments~\cite{Vahsen:2021gnb}. While pixelated readouts provide the most detailed charge topology data~\cite{Vahsen:2014fba,Ligtenberg:2021viw}, past simulation studies suggest that $x/y$ strips can achieve similar 3D directional performance for nuclear recoils, but at a much lower cost~\cite{Vahsen:2020pzb}. TPCs with strip readout are therefore thought to be advantageous for building larger detectors for future neutrino and dark matter experiments. It is thus important for us to experimentally verify that the 3D direction of nuclear tracks can indeed be clearly reconstructed. Potential concerns include the combinatorial ambiguities associated with reconstructing 3D space points from 2D strip hits, and charge spreading effects due to the DLC layer. Here, we develop a method for 3D reconstruction in detectors with digitized $x/y$ strip data. The algorithm is demonstrated on the task of reconstructing alpha particle trajectories, emitted by a radioactive Po-210 source.

VMM/SRS data is recorded with the Po-210 source placed above the cathode for each detector, as shown in Figure~\ref{MM_Po210}. For all analysis involving 3D reconstruction, we only consider the best $x/y$ charge sharing quadrant (quadrant $a$) of the UH detectors. The Po-210 source emits alpha particles that create a straight line of ionization in the sensitive volume. A Micromegas mesh voltage of $V_\textrm{Mesh} = 540$\,V is used to amplify the primary ionization. Similarly to Section~\ref{VMM_Fe55}, the first step in analysis involves clustering raw hits into events using \texttt{VMM-sdat}~\cite{vmm-sdat}; the parameters utilized are detailed in the Po-210 column of Table~\ref{vmmsdat_params1}. A clustered event comprises a collection of timestamped $x$ and $y$ hits. In our notation, ${ x_i }$, ${ e^x_i }$, and ${ t^x_i }$ denote the positions, detected electrons, and timestamps of the $x$ hits within a cluster and ${ y_j }$, ${ e^y_j }$, and ${ t^y_j }$ denote the positions, detected electrons, and timestamps of the $y$ hits.

\begin{figure}[ht]
\begin{center}
\includegraphics[width=0.7\textwidth,trim={.3cm .4cm 0.3cm 0.4cm},clip]{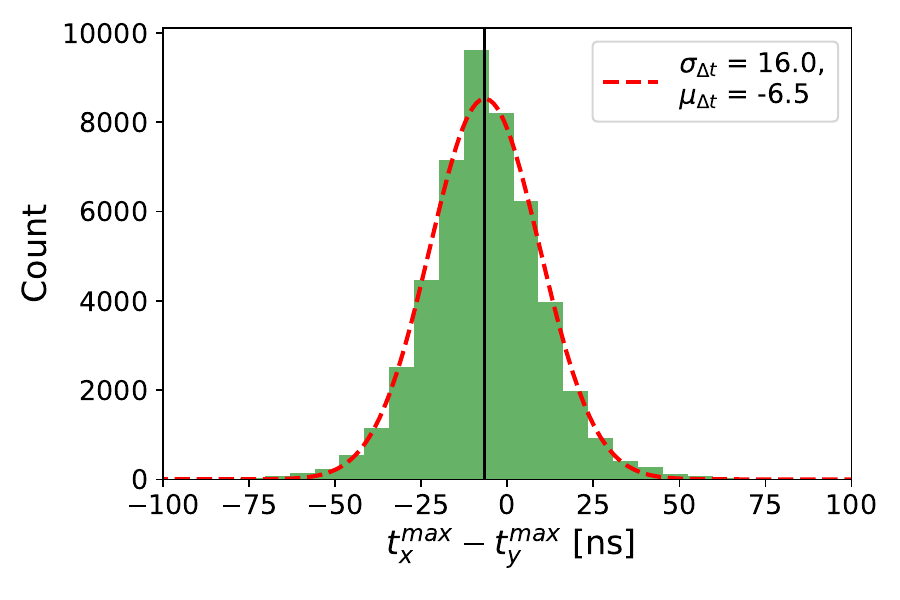}
%\framebox[4.0in]{$\;$}
%\fbox{\rule[-.5cm]{0cm}{4cm} \rule[-.5cm]{4cm}{0cm}}
\end{center}
\caption{ Distribution of the timestamp difference, $\Delta t = t^\textrm{max}_x - t^\textrm{max}_y$, where $t^\textrm{max}_x$ ($t^\textrm{max}_y$) is the timestamp of the $x$ ($y$) hit with the most detected charge for each clustered Fe-55 event, for UH DLC quadrant $a$ with $V_\textrm{Mesh} = 690$\,V. The dashed red line is a Gaussian which has been fit to the distribution to determine $\mu_{\Delta t}$ and $\sigma_{\Delta t}$.}
\label{time_off_dist}
\end{figure}

\begin{figure}[ht]
    \begin{subfigure}{.495\textwidth} 
        \centering
        % include first image
        \includegraphics[width=\linewidth,trim={0.4cm .1cm 0.4cm 0.35cm},clip]{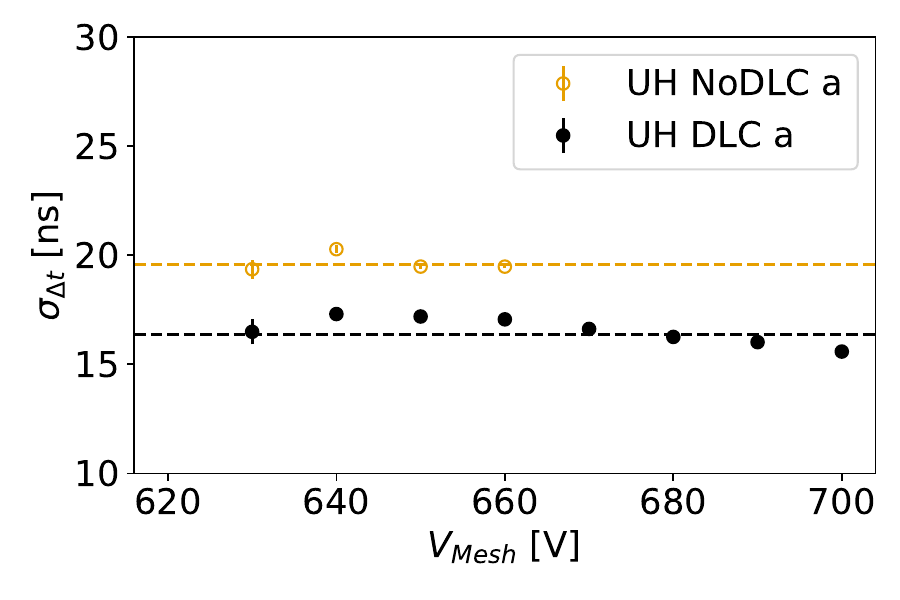}  
        \caption{}
        \label{sigma_delta_t}
    \end{subfigure}
    \begin{subfigure}{.495\textwidth}
        \centering
        % include second image
        \includegraphics[width=\linewidth,trim={0.4cm .1cm 0.4cm 0.35cm},clip]{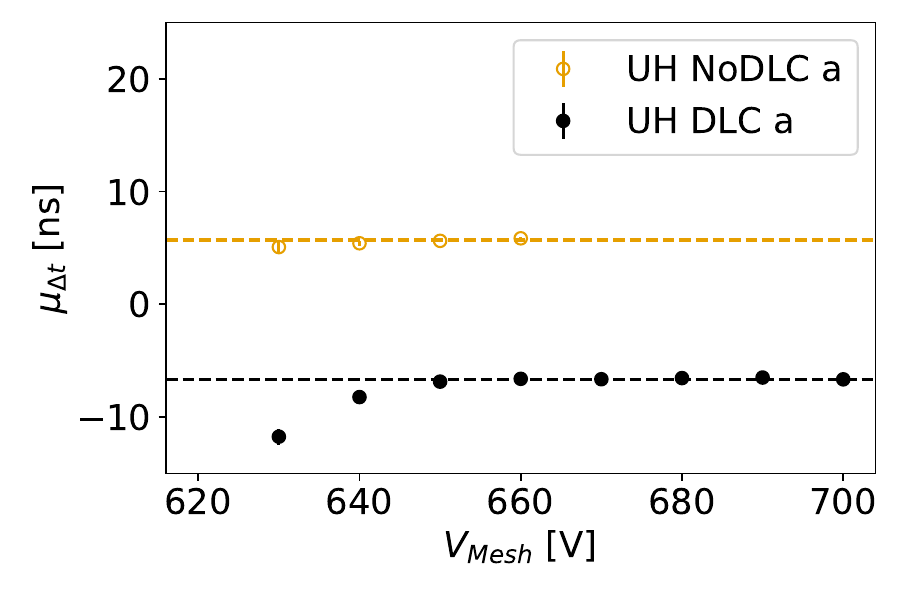}  
        \caption{}
        \label{mu_delta_t}
    \end{subfigure}
    \caption{(a) The standard deviation of the distribution of time differences between the maximum ADC hit in x and y over events versus Micromegas mesh voltage for quadrant $a$ of both UH detectors. (b) The mean of the distribution of time differences between the maximum ADC hit in x and y over events versus Micromegas mesh voltage for quadrant $a$ of both UH detectors. The dashed lines indicate the average over Micromegas mesh voltage for each detector.}
\end{figure}

Our 3D reconstruction algorithm operates on each clustered event by matching the event's $x$ and $y$ hits based on their timestamps, $t^x_i$ and $t^y_j$. To calibrate the timestamp difference between $x$ and $y$ hits corresponding to the same primary ionization, we first revisit the data from Section~\ref{VMM_Fe55}. For each Fe-55 data run, we find a distribution of $\{ t^\textrm{max}_x - t^\textrm{max}_y \}$, where $t^\textrm{max}_x$ ($t^\textrm{max}_y$) is the timestamp of the $x$ ($y$) hit with the most detected charge for each clustered event within the run. An example of this distribution for the $V_\textrm{Mesh} = 690$\,V in the UH DLC detector, quadrant $a$, is displayed in Figure~\ref{time_off_dist}. A Gaussian is fit to each distribution to determine the mean ($\mu_{\Delta t}$) and the standard deviation ($\sigma_{\Delta t}$). These fit values are plotted versus $V_\textrm{Mesh}$ for both UH detectors in Figures~\ref{sigma_delta_t} and~\ref{mu_delta_t}. The average of $\mu_{\Delta t}$ and $\sigma_{\Delta t}$ over $V_\textrm{Mesh}$ for each detector, $\bar{\mu}_{\Delta t}$ and $\bar{\sigma}_{\Delta t}$, is presented as a dashed line. The value of $\bar{\mu}_{\Delta t}$ is $-6.68$\,ns, and $5.70$\,ns for UH DLC quadrant $a$ and UH NoDLC quadrant $a$, respectively. The value of $\bar{\sigma}_{\Delta t}$  is $16.4$\,ns and $19.6$\,ns for UH DLC quadrant $a$ and UH NoDLC quadrant $a$, respectively.

Our 3D reconstruction algorithm uses $\bar{\mu}_{\Delta t}$ and $\bar{\sigma}_{\Delta t}$ to reconstruct clustered events, as follows:
\begin{enumerate}
    \item Find all $i, j$ pairs such that $(t^x_i - t^y_j - \bar{\mu}_{\Delta t})/ \bar{\sigma}_{\Delta t} < 3$. These paired $x/y$ hits are referred to as vertices.
    \item Define $N^x_i$ and $N^y_j$ as the number of vertices in which the ith $x$ hit and jth $y$ hit appear, respectively. Assign a charge $e_{ij} = e^x_i / N^x_i  + e^y_j / N^y_j $ to each vertex. Doing so, each hit's charge is equally shared between the vertices in which it appears.
    \item Assign a timestamp to each vertex calculated by $t_{ij} = (t^x_i + t^y_j)/2 $, the average of the paired hits.
    \item Assign 3D coordinates to the vertices as $(x_i, y_j, z_{ij})$, where $z_{ij} = v_\textrm{drift} t_{ij}$ and $v_\textrm{drift}$ is the drift speed. These coordinates establish the event geometry with absolute (x, y)-coordinates and relative z-coordinates.
    \item Adjust the timestamps of unmatched hits by subtracting (adding) $\bar{\mu}_{\Delta t}/2$ to the timestamp of the $x$ ($y$) hits.
    \item Distribute the charge detected on unmatched hits across all vertices. This distribution is weighted by the inverse of the difference between $t_{ij}$ and the adjusted hit time.
\end{enumerate}
An offline mask is applied to channels 296, 306, 307, 326 which were found to produce irregular hits. Furthermore, x and y hits with a gap of $\geq 2$\,strips to a neighboring hit are identified as noise and removed.

\begin{figure}
\centering
\begin{subfigure}[b]{0.9\textwidth}
   \includegraphics[width=\linewidth,trim={1cm 4.1cm .5cm 5.8cm},clip]{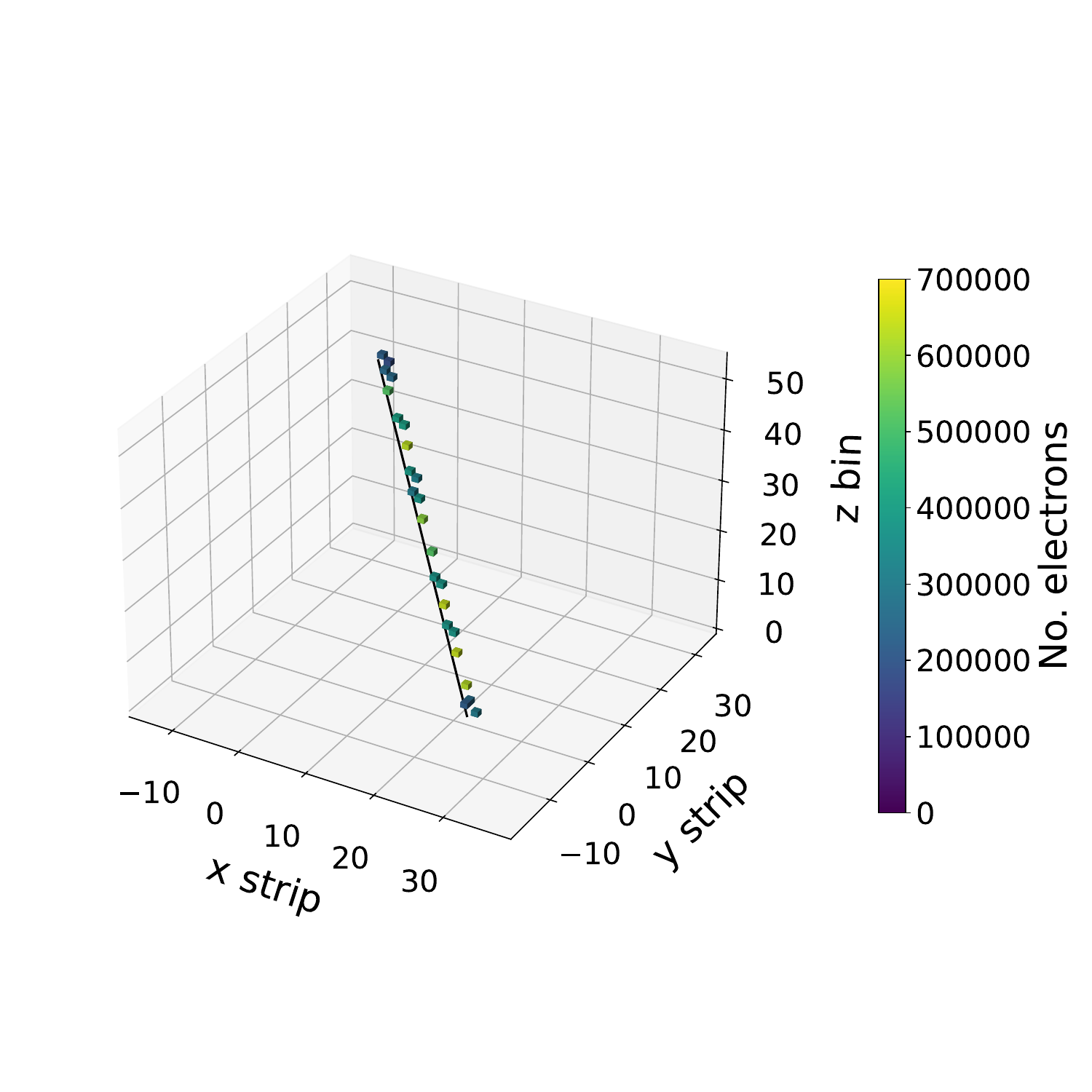} 
   \caption{}
   \label{real_alpha} 
\end{subfigure}

\begin{subfigure}[b]{0.9\textwidth}
   \includegraphics[width=\linewidth,trim={1cm 4.1cm .5cm 5.8cm},clip]{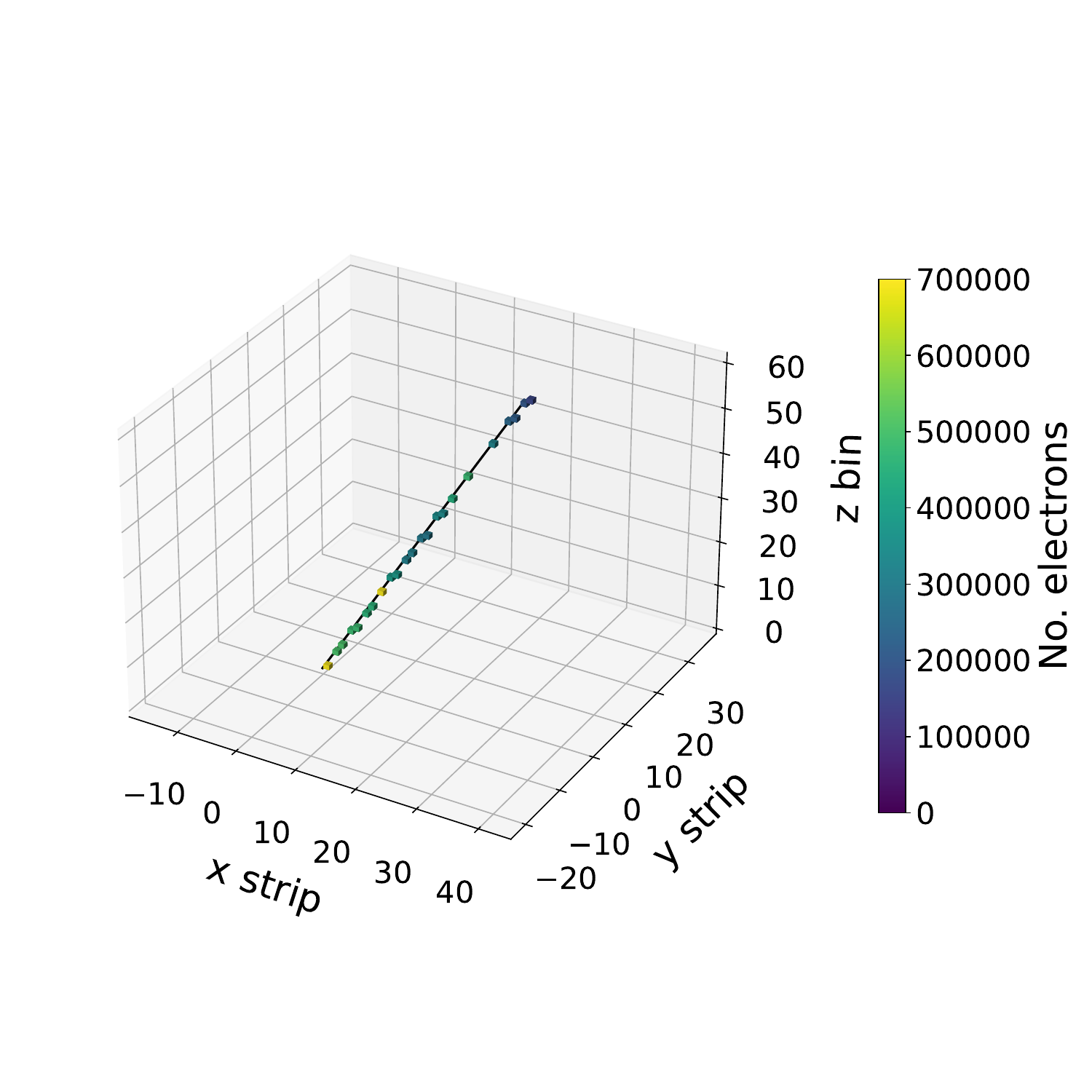}  
   \caption{}
   \label{sim_alpha}
\end{subfigure}

\caption{Po-210 alpha-track events in quadrant $a$ of the UH DLC detector with $V_\textrm{Mesh} = 540$ V for (a) experimental data and (b) simulated data. Each track is reconstructed in 3D, using the algorithm described in the text. The $z$-axis is segmented into 200\,$\upmu \textrm{m}$ bins to match the segmentation of the readout plane.}
\end{figure}

An example of an alpha track reconstructed in 3D using our algorithm on data from the UH DLC detector is presented in Figure~\ref{real_alpha}.
For comparison, the 3D reconstruction algorithm is also applied to a simple simulation of alpha particles in our detectors. The simulation assumes that alpha particles travel in a straight trajectory. \texttt{SRIM}~\cite{srim} is used to determine dE/dx, which is used to simulate energy deposition along the tracks. The alpha-particle direction is drawn isotropically within  an angle $15^\circ < \theta < 30^\circ$ from vertical. Diffusion along the 12\,mm drift length is simulated using the transverse and longitudinal diffusion coefficients, $\sigma_T = 135\, \upmu \textrm{m} / \sqrt{\textrm{cm}}$  and $\sigma_L = 129\, \upmu \textrm{m} / \sqrt{\textrm{cm}}$ , as obtained from \texttt{Magboltz}~\cite{magboltz}. The drift speed is also obtained from \texttt{Magboltz} as $8$\,$\upmu$m/ns. The simulated gain is obtained by substituting $V_\textrm{Mesh} = 540$\,V into Equation~\ref{lin_fit}, which results in a gain of 604 for the UH detectors. After amplification, the avalanche charge is read out by the simulated upper and lower strips independently. The pitch of the strips is 200\,$\upmu$m. The charge is shared between the  upper and lower strips according to the mean $x/y$ charge sharing values in Figures~\ref{VMM_CS_UHDLC}--\ref{VMM_CS_UHNoDLC}. Each strip is assumed to integrate the charge above it for a duration of 200\,ns, equal to the VMM peaking time setting. If the integrated charge exceeds the threshold, as measured in Section~\ref{calib}, a hit is formed. The timestamp of each hit is determined as the charge-weighted mean time of the integrated charge. For consistency with experimental data, each strip is only allowed to form one hit per event, channels 296, 306, 307, and 326 are masked, and hits with a gap of $\geq 2$\,strips to their nearest neighbor identified as noise and removed. An example of a 3D reconstructed alpha particle track based on simulated data is displayed in Figure~\ref{sim_alpha}.

\subsection{Spatial Resolution}
\label{pointres}

\begin{figure}[ht]
    \begin{subfigure}{.49\textwidth} 
        \centering
        % include first image
        \includegraphics[width=\linewidth,trim={0cm .4cm 0.2cm 0cm},clip]{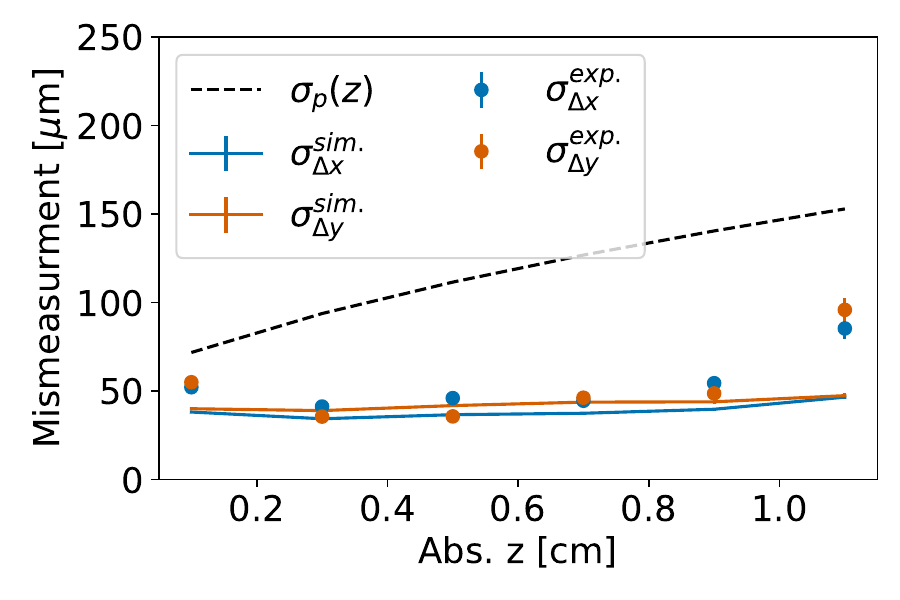}  
        \caption{}
        \label{PR_UHDLC}
    \end{subfigure}
    \begin{subfigure}{.49\textwidth}
        \centering
        % include second image
        \includegraphics[width=\linewidth,trim={0cm .4cm 0.2cm 0cm},clip]{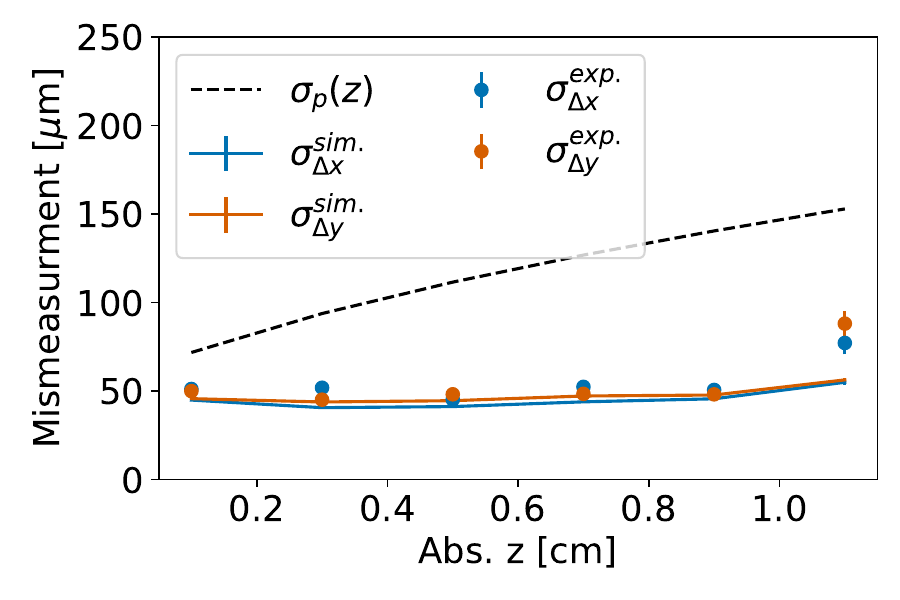}  
        \caption{}
        \label{PR_UHNoDLC}
    \end{subfigure}
    \caption{Position measurement error in $x$ and $y$ versus absolute $z$, for Po-210 alpha particle tracks in simulation and experiment, for a Micromegas mesh voltage of $V_\textrm{Mesh} = 540$\,V. (a) UH DLC (quadrant $a$). (b) UH NoDLC (quadrant $a$). The dashed black line is the expression in Equation~\ref{PR_expected}.}
    \label{VMM_PT}
\end{figure}

We assess the spatial resolution of the UH readout planes with the Po-210 data described in Section~\ref{3Drecon} by following a methodology analogous to that outlined in Ref.~\cite{Vahsen:2014fba}. Because the VMM channels are operating in a self-triggering mode, there is an ambiguity in the absolute $z$ position for individual hits. We circumvent this by selecting, both in simulation and experiment, tracks that traverse the entire drift length and are within an angle $15^\circ < \theta < 30^\circ$ from vertical. The absolute $z$ position of hits is then obtained by identifying the reconstructed event's lowest 3D vertex as $z=0$. Only reconstructions with over five vertices are considered for this analysis. The principle axis of each alpha track is obtained using SVD and the measurement error for each reconstructed vertex is quantified as the sign $1D$ distance, in $x$ and $y$, from the vertex to the principle axis. The $x$ and $y$ measurement errors of all vertices in all tracks, are binned with respect to the absolute $z$ position of their vertex. For each bin in absolute $z$, the measurement errors in $x$ and $y$ are fitted to a Gaussian distribution to determine $\sigma_{\Delta x}$ and $\sigma_{\Delta y}$, respectively. To distinguish between simulated and experimental results, we append a superscript (either sim. or exp.) to $\sigma_{\Delta x}$ and $\sigma_{\Delta y}$. Figure~\ref{VMM_PT} illustrates $\sigma_{\Delta x}$ and $\sigma_{\Delta y}$ versus absolute $z$ for both simulation and experiment.

The simulated and measured spatial resolutions agree fairly well. It is, however, instructive to also compare these against a naive, analytical prediction of the spatial resolution, illustrated by the dashed black line in Figure~\ref{VMM_PT}, and given by
\begin{equation}
    \sigma_x(z) = \sigma_y(z) = \sqrt{(200\,\upmu\textrm{m}/\sqrt{12})^2 + (\sigma_T \sqrt{z} )^2 },
    \label{PR_expected}
\end{equation}
where the first term on the right-hand side is the expected spatial resolution due to readout segmentation and the second term is the transverse diffusion contribution. While a similar prediction worked fairly well for pixel ASIC charge readout~\cite{Vahsen:2014fba}, it is naive, as it does not consider any effects due to signal induction in the strips, front end electronics, or strip to vertex conversion ambiguities. One might generally expect that the naive estimate would be an under-estimate, due to excluding certain smearing effect. However, for the $x/y$ strip readouts tested here, we find the opposite: in both Figures~\ref{PR_UHDLC} and~\ref{PR_UHNoDLC}, $\sigma_{x/y}(z)$ is larger than the measured $\sigma_{\Delta x}$ and $\sigma_{\Delta y}$ for simulation and experiment. By re-running the simulation with different effects turned on and off, we are able to identify that there are multiple effects that suppress the observed measurement error in simulation, and therefore likely are also the cause for this in experimental data, which matches the simulation reasonably well. For the lowest values of absolute $z$, our simulations suggest that the discrepancy between $\sigma_p(z)$ and simulation/experiment is due to the threshold. In the following bins, the diffusion appears to be suppressed in both experiment and simulation. This effect is due to the use of charge integrating strips that require time to digitize data before they can produce another hit. For the highest-$z$ bin, $\sigma_{\Delta x}$ and $\sigma_{\Delta y}$ are larger in experiment than in simulation for both Figures~\ref{PR_UHDLC} and~\ref{PR_UHNoDLC}. This discrepancy could be due to the simulation parameters. In simulation, each strip integrates charge for 200\,ns, the VMM peaking time. However, if the strip integration time is reduced to 100\,ns, the agreement between simulation and experiment is improved. Another possible explanation is that the increased $\sigma_{\Delta x}$ and $\sigma_{\Delta y}$ in experiment is caused by an interaction between the alpha tracks and the cathode mesh, which is not accounted for in simulation. 

To quantify the agreement between simulation and experiment in Figures~\ref{PR_UHDLC} and~\ref{PR_UHNoDLC}, the data is fitted to
 \begin{equation}
\sigma^\textrm{exp.}_{\Delta x/y} = \sqrt{ (\sigma^\textrm{sim.}_{\Delta x/y})^2 + (\sigma^o_{x/y})^2 }, 
\label{PR_fit1}
 \end{equation}
where $x/y$ indicates that the subscript is either $x$ or $y$. Here, $\sigma^o_{x/y}$ [$\upmu$m] is a fit parameter. Several effects can contribute to the spatial resolution that are not included in the simulation, such as: charge spreading in the amplification gap, electric field non-uniformity, straggling of the alpha particles, and spreading in the induced signal on the upper and lower strips. However, contributions to the spatial resolution can be suppressed in an $x/y$ strip readout, as illustrated in Figure~\ref{VMM_PT}. Therefore, $\sigma^o_{x/y}$ is interpreted as a measure of the agreement between simulation and experiment, rather than a constraint on the spatial resolution contributions that are not accounted for in simulation. In fitting the data, we omit the point corresponding to the highest bin in absolute z as it could be an artifact that is not accounted for in simulation. The results are summarized in Table~\ref{mismeasurments}. We observe similar performance between the UH DLC and NoDLC detectors with $\sigma^o_{x/y}$ being small with respect to the 200\,$\upmu$m strip pitch of the detectors. In both cases $\sigma^o_y$ is smaller than $\sigma^o_x$. This could be because the $x$ strips are lower than the $y$ strips and our simple simulation does not model the detailed signal induction. 

\begin{table}[h!]
\centering
\begin{tabular}{|c| c|c |} 
 \hline
 Detector & $\sigma^o_x$ [$\upmu$m] & $\sigma^o_y$  [$\upmu$m] \\ [0.5ex] 
 \hline
 UH DLC Quadrant $a$ & $29.4 \pm 2.24$ & $10.3 \pm 4.98$   \\ 
 UH NoDLC Quadrant $a$ & $25.6 \pm 2.63$ & $14.3 \pm 4.55$ \\ 
 \hline
\end{tabular}
\caption{Values of the $\sigma^o_{x/y}$ fit parameter from Equation~\ref{PR_fit1}. These represent a measure of how closely the simulated and experimental spatial resolution agree.}
\label{mismeasurments}
\end{table}

\section{Summary and Discussion of Results}
\label{Discussion}

\subsection{This Work}

By comparing the UH DLC and NoDLC detectors in Sections~\ref{PHA} and~\ref{VMM}, we find that $x/y$ strip charge readouts have similar gain versus Micromegas mesh voltage response, regardless of whether or not they utilize a DLC layer. However, the DLC layer provides spark protection, allowing UH DLC to reach higher Micromegas mesh voltages and therefore larger gain. The highest gain achieved by UH DLC is $16.2\times10^3$, a factor of 2.29 larger than the highest gain achieved by the UH NoDLC detector, $7.06 \times 10^3$. The UoS detector which featured a DLC layer and a larger Micromegas amplification gap achieved the highest overall gain, $76.8 \times 10^3$. The positional dependence of the avalanche gain is tested in Section~\ref{PHA}, Table~\ref{quadrants} by moving the Fe-55 source around the readout and measuring the avalanche gain using the PHA system on the Micromegas mesh. A positional variation of 4\%, 5\%, and 6\% is noted for the UoS, UH DLC, and UH NoDLC detectors, respectively. In Section~\ref{VMM}, we see a larger positional variation in the effective gain, as measured by the VMM/SRS system on the strips, in the UH detectors which have quadrants with different upper strip width. The effective gain measured on UH DLC quadrant $d$ (thickest upper strips) is 21.9\% larger than quadrant $a$ (thinnest upper strips). Similarly, for the UH NoDLC detector, the effective gain was 15.7\% larger in quadrant $d$ than $a$. These findings show that the strip configuration has little impact on the avalanche gain, as measured by the PHA system on the Micromegas mesh, but a larger impact on the effective gain as measured by VMM/SRS on the strips. In general, configurations with wider upper strips observe a larger effective gain. A likely explanation for the observed results is that the strip configuration has little impact on the amplification field strength; however, configurations with more conductor within proximity of the charge avalanche see a larger induced signal on the strips.

Another benefit of the DLC layer is improved fractional gain resolution. This is likely because the inclusion of a grounded DLC layer makes the electric field below the Micromegas mesh more uniform. In Section~\ref{PHA}, we found that the asymptotic value of the fractional avalanche gain resolution is $0.922 \pm 0.001$ and $0.093 \pm 0.001$ for the UH DLC and UoS detectors which had a DLC resistivity of $70$ and $50$\,M$\Omega$/sq, respectively. The UH NoDLC detector had a comparatively worse asymptotic fractional gain resolution of $0.120 \pm 0.002$. Furthermore, in Figure~\ref{VMM_GainRes_UH} we see that the UH DLC detector's fractional effective gain resolution, as measured with VMM/SRS, approaches the asymptotic value of the fractional avalanche gain resolution measured in Section~\ref{PHA}. This indicates that the strip readout contribution to the fractional gain resolution in negligibly small with respect to Micromegas contribution.

In Section~\ref{VMM_Fe55}, we see that both the DLC layer and strip configuration affect the $x/y$ charge sharing. For example, the mean $x/y$ charge sharing is 4.38 and 3.32 times larger in quadrant $a$ than in quadrant $d$ for the UH DLC and NoDLC detectors, respectively. This suggests that reducing the upper strip width significantly increases the $x/y$ charge sharing. Furthermore, in quadrant $a$, UH NoDLC has a mean $x/y$ charge sharing value that is 51\% larger than UH DLC. Hence the inclusion of the DLC layer decreases the $x/y$ charge sharing. For both UH detectors, thinner upper strips are required if 50/50 charge sharing is desired.

In Sections~\ref{3Drecon} and~\ref{pointres}, alpha tracks emitted from a Po-210 source were reconstructed to assess the spatial resolution of UH detectors in quadrant $a$. In both cases we found good agreement between experimental data and simulations. In Table~\ref{mismeasurments} we quantify the extent to which experimental data agree with simulations, finding UH DLC and NoDLC perform similarly.

Considering all results, including a DLC layer appears to outweigh any drawbacks: A DLC layer allows higher Micromegas mesh voltage before sparking occurs, protects VMM channels from damage when sparking does occur, improves the fractional gain resolution, and has no notable impact on the spatial resolution. While the DLC layer reduces the $x/y$ charge sharing, this can be mitigated via configurations with thinner upper strips.

\subsection{Future Directions}

The 3D reconstruction algorithm, detailed in Section~\ref{3Drecon}, relies on matching $x$ and $y$ hits based on their timestamps to form vertices. However, this process is prone to combinatorial ambiguities, present in all $x/y$ strip readouts. For instance, if an ionization track runs parallel to the readout plane, all of the hits will share the same timestamp, making it impossible to reconstruct the track accurately. Furthermore, Section~\ref{pointres} indicates that the use of charge integrating strips with a digitization time artificially suppresses the apparent diffusion width of the reconstructed tracks. This is an unexpected finding, and it will be important to evaluate in future work how this impacts particle identification capabilities, fiducialization, and angular resolution for low-energy nuclear recoils. Such work will involve characterization of 3D reconstruction capabilities at much higher gain than reported here. Fully understanding and optimizing the detector response in that regime is likely to require a detailed simulation of charge induction in the strips. 

Both of these issues (combinatorial ambiguities and suppression of apparent diffusion) are due to the fact that an $x/y$ strip readout is being used to integrate a (amplified) 3D charge track. Electron counting is a proposed solution that could resolve both issues while also significantly improving the energy resolution of the detector. In electron counting, a negative ion drift (NID) gas~\cite{MARTOFF2000355,Snowden-Ifft:2014taa,Phan:2016veo} would be used to reduce drift speeds such that avalanche pulses created by individual primary electrons can be resolved and counted. This technique would allow for the primary ionization to be counted directly, thereby ensuring that gain fluctuations do not affect the energy resolution. The ability to resolve individual electrons removes ambiguities when matching $x$ and $y$ hits and resolves the diffusion-suppressing effects discussed in Section~\ref{pointres}. The use of NID gas also reduces diffusion, thereby enhancing the detector's position resolution. Electron counting would be a breakthrough in the field of directional recoil detection. To date, the most advanced demonstration reached an electron detection efficiency of 78\%, as reported in Ref.~\cite{SORENSEN2012106} and the energy resolution did not behave as expected.

In Section~\ref{Noise}, the noise level of the UH detectors is measured as $\sigma_\textrm{noise} \approx 1500$\,electrons. For a well-grounded and shielded detector, we anticipate that threshold can be set $6 \, \sigma_\textrm{noise}$ above pedestal, so that 9000\,electrons are needed to trigger a hit. We assume that the avalanche for a single primary electron is contained above a single 200\,$\upmu$m strip in $x$ and $y$. The mean x/y charge sharing of the UH DLC detector quadrant a is 0.41, meaning that the lower (upper) strips observe 29\% (71\%) of the amplified charge. With this x/y charge sharing, 31000 (12700) avalanche electrons would be required to produce an above threshold signal on the lower (upper) strips. We further assume that the avalanche gain for a single primary electron is drawn from an exponential distribution
\begin{equation}
        f_{exp}(x)=
        \begin{cases}
            0 , & \textrm{for } x < 0 \\
            \frac{1}{\mu_e} \exp{[\frac{-x}{\mu_e}]}, & \textrm{for } x \geq 0
        \end{cases},
\label{exponential}
\end{equation}
where $\mu_e$ is the mean value of the gain. Let us define electron counting as the point were 80\% of the primary electrons are detected and counted. The expected fractional energy resolution 5.2\%~\cite{SORENSEN2012106}, a substantial improvement over the asymptotic fractional gain resolution measured is sections~\ref{PHA} and~\ref{VMM}. With this definition, in order to achieve electron counting in the lower strips, the required (mean) gain is given by solving
\begin{equation*}
    0.20 = \int_0^{31000} \frac{1}{\mu_e} \exp{[\frac{-x}{\mu_e}]} dx,
\end{equation*}
to obtain $\mu_e = 1.39 \times 10^5$. If electron counting is only desired on the upper strips, then $\mu_e = 5.68 \times 10^4$. In Section~\ref{PHA}, the maximum gain of $7.73 \times 10^4$ was achieved by the UoS detector which has a 256\,$\upmu$m amplification gap. Although these gains were attained with an electron drift gas mixture, similar gains have been demonstrated using a novel multi-mesh ThGEM amplification structure in NID gas mixtures as detailed in Ref.~\cite{McLean:2023dnh}. While the multi-mesh ThGEM achieves the desired gain, its granularity is too coarse. This suggests electron counting with negative ion drift and $x/y$ strip readouts might require an improved MPGD amplification device. The alternative, electron counting without NID, would require re-optimized readout electronics, capable of faster refresh rates while maintaining low noise. 

\section{Conclusion}
\label{concolusion}
We have compared nine unique $x/y$ strip readout configurations coupled to a Micromegas amplification structure. For these comparisons, we instrumented the Micromegas mesh with a PHA setup and the strips with RD51 VMM/SRS electronics. This comparative analysis demonstrated that the DLC layer not only enables detectors to achieve higher gains but also improves their fractional gain resolution. Conversely, the DLC layer tends to reduce the $x/y$ charge sharing. We have also tested the relationship between the width of the upper strips and the $x/y$ charge sharing, with and without the DLC layer. Configuration with thinner upper strips observe an improved charge sharing but at the cost of a slightly decreased effective gain. Additionally, we have developed an algorithm to reconstruct digitized VMM data in three dimensions. Utilizing this algorithm, we find that the DLC layer does not have a noticeable effect on the spatial resolution of a detector. Our results are valuable in informing the design of a future $x/y$ strip readout for the next generation of detectors we a currently building. Our findings also suggest that electron counting in $x/y$ strip readouts could be achievable in the near future, but it will require an amplification device that can reach the gain of the multi-mesh ThGEM with improved granularity. Future work will involve a more detailed study of the electronic noise with a more stable noise scan and improved grounding scheme. A more detailed simulation that includes charge induction on the strips in also left for future work.

\section*{Acknowledgements}
\noindent The authors thank Rui De Oliveira and Bertrand Mehl for manufacturing the detectors and for useful discussions on the design of the readouts. MG expresses gratitude to Florian Maximilian Brunbauer, Mauro Iodice, and all organizers of the RD51 MPGD school. We gratefully acknowledge helpful discussion with Lucian Scharenberg on the VMM/SRS system. MG and SEV acknowledge support from the U.S. Department of Energy (DOE) via Award Number DE-SC0010504. AGM and NJCS would like to acknowledge the ``University of Sheffield EPSRC Doctoral Training Partnership (DTP) Case Conversion Scholarship'' awarded to AGM. FD acknowledges the support of the ARC Centre of Excellence for Dark Matter Particle Physics.



\bibliographystyle{elsarticle-num} 
\bibliography{mybib}

\end{document}